\RequirePackage{fix-cm, amsmath}
\documentclass[smallextended]{svjour3}       
\smartqed  
\usepackage{amsmath}
\usepackage{graphicx}
\usepackage{cite}
\usepackage{amssymb,amsfonts}
\usepackage{algorithmic}
\usepackage{textcomp}
\usepackage{xcolor}
\usepackage{tcolorbox}
\usepackage{color, colortbl}
\usepackage{flushend}
\usepackage{multirow}
\usepackage{url}

\def\BibTeX{{\rm B\kern-.05em{\sc i\kern-.025em b}\kern-.08em
    T\kern-.1667em\lower.7ex\hbox{E}\kern-.125emX}}

\definecolor{gray}{rgb}{0.5,0.5,0.5}

\newcommand{\eg}{{\textit{e.g.,}}}
\newcommand{\ie}{{\textit{i.e.,}}}
\newcommand{\etal}{\textit{et al.}}

\begin{document}

\title{Machine Learning Application Development: Practitioners' Insights
}


\author{Md Saidur Rahman \and
        Foutse Khomh \and
        Alaleh Hamidi \and
        Jinghui Cheng \and
        Giuliano Antoniol \and
        Hironori Washizaki
        }


\institute{Md Saidur Rahman (Corresponding author) \at
           	SWAT Lab, DGIGL, Polytechnique Montr\'{e}al, Montr\'{e}al, QC, Canada\\
           	\email{saidur.rahman@polymtl.ca}
           	 \and
            Foutse Khomh 
            \at	SWAT Lab, DGIGL, Polytechnique Montr\'{e}al, Montr\'{e}al, QC, Canada\\
           	\email{foutse.khomh@polymtl.ca}
           	\and
           Alaleh Hamidi \at
           	SWAT Lab, DGIGL, Polytechnique Montr\'{e}al, Montr\'{e}al, QC, Canada\\
           	\email{alaleh.hamidi@polymtl.ca}
      \and 
            Jinghui Cheng \at
           	DGIGL, Polytechnique Montr\'{e}al, Montr\'{e}al, QC, Canada\\
           	\email{jinghui.cheng@polymtl.ca}
      \and 
            Giuliano Antoniol \at
           	DGIGL, Polytechnique Montr\'{e}al, Montr\'{e}al, QC, Canada\\
           	\email{giuliano.antoniol@polymtl.ca}
      \and 
           Hironori Washizaki \at
           	Waseda University, Japan\\
           	 \email{washizaki@waseda.jp}
}

\date{Received: date / Accepted: date}

\maketitle

\begin{abstract}
Nowadays, intelligent systems and services are getting increasingly popular as they provide data-driven solutions to diverse real-world problems, thanks to recent breakthroughs in Artificial Intelligence (AI) and Machine Learning (ML). 
However, machine learning meets software engineering not only with promising potentials but also with some inherent challenges. 
Despite some recent research efforts, we still do not have a clear understanding of the challenges of developing ML-based applications and the current industry practices. Moreover, it is unclear where software engineering researchers should focus their efforts to better support ML application developers. 
In this paper, we report about a survey 
that aimed to understand the 
challenges and best practices of ML application development. 
We synthesize the results obtained from 80 practitioners (with diverse skills, experience, and application domains) into 17 findings; 
outlining challenges and best practices for ML application development. Practitioners involved in the development of ML-based software systems can leverage the summarized best practices to improve the quality of their system. We hope that the reported challenges will inform the research community about topics that need to be investigated to improve the engineering process and the quality of ML-based applications. 
\end{abstract}

\section{Introduction}
\label{intro}
Artificial Intelligence (AI) and Machine Learning (ML) have emerged as powerful tools to develop data-driven solutions for diverse real-world problems. 
Recent breakthroughs in machine learning have greatly inspired the surging adoption of AI capabilities for automation by embedding intelligence into modern software and services \cite{Amershi_ICSE_2019}. 
AI-based automated supports now span almost every sphere of human life: business, education, healthcare, research, communication, security, assistive technologies and so on. 
With the diversity in application domains, the types of problems and the characteristics of the data may vary greatly and so the ML algorithms. From an engineering perspective, once an algorithm is implemented, it requires a solid architecture, model/data validation, proper monitoring for changes, dedicated release engineering strategies, judicious adoption of design patterns and security checks, and thorough user experience evaluation and adjustment. A failure to properly address these challenges can lead to catastrophic consequences. Classically, we have constructed software systems in a deductive way, or by writing down the rules that govern the system behaviors as program code. With machine learning techniques, we generate such rules in an inductive way from training data. This shift of paradigm induces some challenges that are unique to ML application development \cite{Foutse_Redhat_2018,Foutse_IEEESW_2019}. 
%

Recently, practitioners from leading software companies like Google \cite{Sculley_NIPS_2015} and Microsoft \cite{Amershi_ICSE_2019} have been reporting about their experience building ML-based applications and raising awareness on some of the challenges posed by %
ML application development. 
Sculley \etal \cite{Sculley_NIPS_2015} outlined some challenges of ML application development by identifying harmful design patterns that may incur excessive maintenance costs. 
In addition to characterizing the challenges, they also made some suggestions on how to deal with those challenges. 
Amershi \etal \cite{Amershi_ICSE_2019} presented a survey conducted with developers from Microsoft, showing how AI application development aligns with a nine-stage development workflow. 
They outlined three fundamental differences between ML application development and traditional software development. 
They observed that \textit{data management} for ML applications is quite complex compared to other types of software, and that \textit{model customization and reuse} requires some specific skills. 
They also reported that AI modules are difficult to handle compared to traditional software components due to \textit{complex inter-component relationships and non-monotonic error} behaviour.
Amershi \etal \cite{Amershi_ICSE_2019} also suggested some best practices for software engineering of ML applications, focusing on data and model management, and the interfaces between ML components and the overall system. 

\begin{figure*}[t]
\center
\includegraphics[scale=0.6]{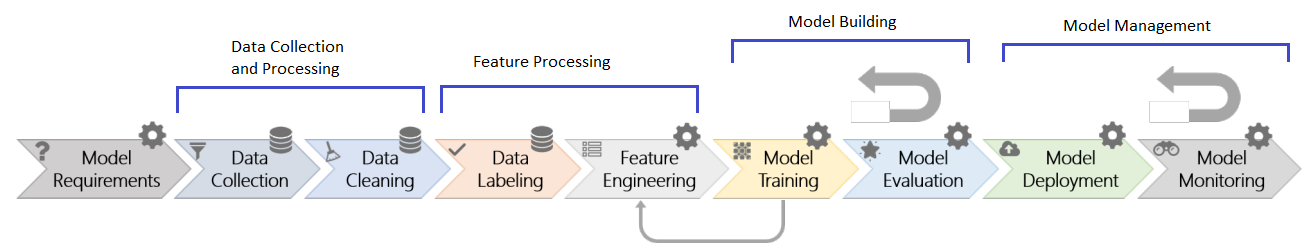}
\caption{Phases of ML workflow (adopted from Amershi \etal \cite{Amershi_ICSE_2019})}
\label{fig:ML-phases}
\end{figure*}

Although these studies (i.e., \cite{Amershi_ICSE_2019}, \cite{Sculley_NIPS_2015}) have provided valuable insights on the challenges of developing AI/ML applications at scale in the context of large companies, we still don't know how small and medium-sized enterprises (SMEs) handle ML application development. It is important to know the challenges and best practices followed by practitioners building ML applications across different domains and in diverse development settings. 
This paper aims to fill this gap by examining experiences and collect insights from ML practitioners from across the globe with varying skills and experiences and from diverse development domains. We present a survey of ML development practices 
and insights obtained from the feedback of 80 ML practitioners working in the software industry or in academia.\\  
For the survey, we reached out to over 700 AI/ML practitioners by email. We communicated our request for participation in the survey using contacts from the professional network LinkedIn. We selected the participants based on their profile information indicating their roles associated with AI/ML in academia or industry. We also collected the emails of the participants from GitHub based on their contributions to ML projects. We received responses from 80 participants with diverse technical and professional background. 
We analyze the survey data to derive insights and summarize them along the phases of the ML development workflow described in \cite{Amershi_ICSE_2019}. 
%

In this paper, we make the following contributions:
\begin{itemize}
    \item We conduct a comprehensive 
    survey involving 80 ML practitioners from diverse backgrounds to identify the state of practices and challenges in ML application development. 
    \item Our survey covers four key phases of ML application development life cycle, namely (1) data collection and preprocessing, (2) feature engineering, (3) model building and testing, and (4) integration, deployment and monitoring, to identify challenges and practices from practitioners' perspective.
    \item We synthesize our 17 key findings to show
    how those findings can benefit researchers and practitioners in developing ML applications of high quality. 
\end{itemize}
Practitioners embarking on new or ongoing efforts to develop ML-based applications can take advantage of the summarized best practices to improve the quality of these applications.

\textbf{The remainder of the paper is organized as follows.} Section \ref{sec:background} discusses some basic concepts of ML application development, common trends in ML application, their benefits and challenges. Section \ref{sec:studysetup} presents the detail of the survey including design, objective, participants, data collection and analysis methodologies. Section \ref{sec:results} presents the results of our survey. 
Section \ref{sec:discussion} discusses these 
results. In Section \ref{sec:threats}, we discuss potential threats to our methodology and findings. Section \ref{sec:relatedworks} presents some prior research related to our study followed by the conclusions in Section \ref{sec:conclusion}.






\section{Background} \label{sec:background}

This section briefly presents some important concepts of ML application development. 
We also briefly compare and contrast traditional software systems and ML-based systems. 

\subsection{Machine Learning Applications}
Traditional software systems are constructed based on a well-defined set of rules that govern the system's behaviour. However, in ML applications the behaviour is 
controlled by rules inferred from the data \cite{Foutse_IEEESW_2019}. 
ML applications as data-driven systems have induced 
a paradigm shift in the software development process, making the development, testing and verification of the ML applications intrinsically harder. A defect in a ML application may come from training data, program code, execution environment, or third-party frameworks. 
Given the increasing adoption of ML/AI, it is important to understand the challenges of ML application development and devise some best practices. Since ML/AI is an emerging field, we believe that developers who are currently building ML applications are best positioned to reflect and report about the challenges and pitfalls of ML application development. Hence, in this paper, we conduct a survey of ML developers to document their experiences and formulate best practices and the challenges of ML application development. 


\subsection{ML Application Development Life Cycle}
In our study, we consider the ML application development life-cycle presented by Amershi \etal \cite{Amershi_ICSE_2019} as shown in Fig. \ref{fig:ML-phases}. We study practitioners' perceptions of the challenges and common practices in ML application development. We briefly discuss the phases of the ML application development life cycle bellow. A more detailed discussion of the ML application development life cycle is available in \cite{HoussemBenBraiek_2018}.

\subsubsection{Model Requirements}

In this phase, developers define the requirements for data and algorithms regarding a ML problem at hand. They need to identify relevant and representative data. %
The requirement is very important since it has a significant impact on the success of the other phases of the ML workflow. Selecting insufficient or biased data will likely lead to inadequate ML models. In this phase, developers also often have to mediate between different conflicting goals. For example, ensuring high performance of models while satisfying restrictions enforced by regulations governing privacy and security of information (which often restrict access to some data). Regulations can also induce requirements on the models. For example the General Data Protection Regulation (GDPR) enforces the right to explanation, which requires that ML models be explainable and interpretable. 
%

\subsubsection{Data Collection and Preprocessing}
ML applications are data-driven and thus the collection and preprocessing of the data is important. In this phase, data is collected from internal or external sources (e.g., mainframe databases, sensors, IoT devices, and software systems) and is presented in different formats (e.g., various media types). It can be structured (such as database records) or unstructured (such as raw text) and is delivered to ML models either in batch (e.g., discrete chunks from mainframe databases and file systems) and/or real-time (e.g., continuous flow from IoT devices or Stream REST APIs). Developers often have to leverage complementary automated tools that support batch and/or real-time data ingestion strategies, to collect data needed for training their ML models. Once data is collected, it often must be cleaned to ensure consistency and the absence of redundancies. 
Common data cleaning tasks include: removing invalid or undefined values (i.e., Not-a-Number, Not-Available), duplicate rows, and outliers that seems to be too different from the mean value); and unifying the variables’ representations to avoid multiple data formats and mixed numerical scales. This preprocessing step is often done using data transformations such as normalization, min-max scaling, and data format conversion.
%


\subsubsection{Feature Engineering} 
Feature engineering is the process of extracting informative features from the data that ML algorithms can learn from to build ML models. Features need to be able to represent the characteristics or patterns in the dataset. Once suitable features are extracted, it is also important to select the best subset of features for the models. This process is called feature selection. Extraction and selection of features comprise the feature engineering process. It is an essential step in the construction of conventional ML models.
However, in the case of deep learning models, the features are inferred automatically. In fact, deep learning models build complex features automatically as a part of their statistical learning process from data. For example, conventional computer-vision models require image features, including edges, corners and blobs that can be detected using low-level image processing operations, while Convolutional Neural Networks process raw images directly.

\subsubsection{Model Training and Evaluation}
During the training phase, a suitable machine learning algorithm is applied to the cleaned and prepared dataset. Different model parameters are tuned iteratively to learn the mapping between the features and the corresponding labels (in case of supervised learning). Models are trained up to a desired level of accuracy. The trained model is evaluated on the validation data set, to evaluate the performance. The performance of the model is measured using a predefined set of performance metrics such as prediction or classification \textit{accuracy}.   

\subsubsection{Integration, Deployment and Monitoring}
Once a trained and validated model is available, it is integrated into the target application for the desired functions. The application is deployed on suitable devices or platforms. Deployed ML models need monitoring for performance and potential errors during real-world executions. 

In case of errors or major shifts in the patterns in the data, the models may need to be 
retrained. Thus, the phases of the ML workflow are not linear as it looks like in Fig. \ref{fig:ML-phases}, rather the phases in the ML application development life cycle are iterative. 

In our study, we focus on the following four phases of ML workflow except the requirements phase namely: data collection and preprocessing, feature engineering, model training and evaluation, model management (covering integration), and model deployment and post-deployment monitoring. 
We do not cover the requirement engineering phase in this survey and we plan a future study of its own. 
This is because requirements engineering for ML is quite complex \cite{Belani_2019, Vogelsang_2019}. 
ML engineering introduces a
paradigm shift compared to conventional software engineering \cite{Wan_IEEETrans_2019} and so the requirements engineering \cite{Vogelsang_2019}. 
ML applications are likely to have ML and non-ML requirements. ML application are often developed as a component interacting with other non-ML components to build large and complex systems. Functional and nonfunctional requirements, ML-specific quality trade-offs, and ML and non-ML components' interactions require different considerations. These  make the requirements engineering of ML application a challenging task.  
Ishikawa and Yoshioka \cite{Ishikawa_2019} in their recent study listed requirements engineering as the most difficult activity for the development of ML systems. 
Our survey thus focus on the above mentioned four phases of ML workflow and identifies the common practices and key challenges in the ML workflow. 


\section{Study Design} \label{sec:studysetup}
\begin{figure*}[t]
\center
\includegraphics[scale=0.65]{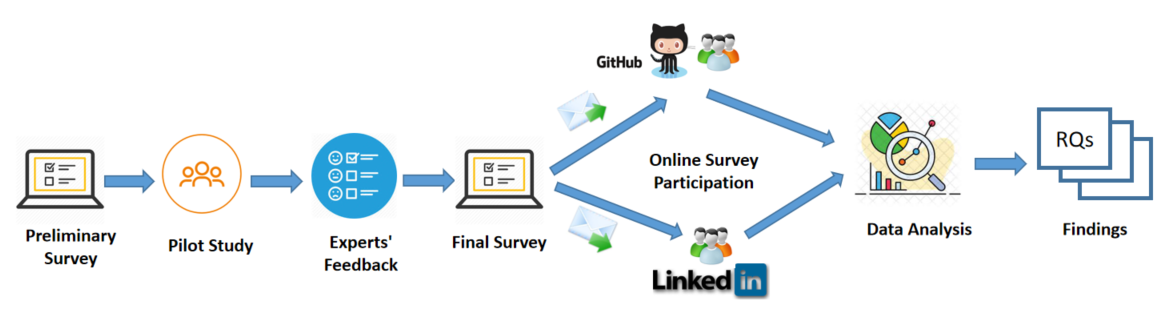}
\caption{Schematic Diagram of the Study}
\label{fig:study}
\end{figure*}

We conducted an online survey to understand the practitioners' experiences in ML application development. We present the overall approach of the study in Fig. \ref{fig:study}. We briefly discuss our study objectives and methodology as follows: 

\subsection{Objectives of the study}
Our key objective in this research is to know the perceptions of the ML practitioners about the challenges and state of practices in developing machine learning applications. 
Using an online survey we ask the developers questions on development activities encompassing different phases of the ML application development life cycle. Our key focus in this study is understanding the challenges and best practices in data collection and preprocessing, feature engineering, ML model building, testing, and deployment. As ML applications are data-driven, we first focus on data processing and feature engineering. We aim to know about the current practices in data processing and feature engineering including source and types of data, data preprocessing activities, tools and frameworks. Then we focus on identifying the challenges and best practices in model building, testing, deployment, and post-deployment model maintenance. 

     
     
     
     
 

 \begin{table*}
\centering
\caption{Research Questions} 
\label{tab:Rq}
{\renewcommand{\arraystretch}{1.2}
\begin{tabular}{p{3cm}|p{7.5cm}}
 \hline
 \textbf{Contexts}   & \textbf{Research Questions}\\   \hline
 \multirow{2}{*}{ML Trends}
    &  \textbf{RQ1:} What are the current industry trends in developing ML applications? \\ \hline
 \multirow{10}{*}{Data Processing} 
     &  \textbf{RQ2:} In practitioner's perception, what are the important quality attributes of ML data?\\ \cline{2-2}
     &  \textbf{RQ3:} What is the state-of-the-practice regarding the data processing tasks, techniques and tools for quality assurance of ML data? \\ \cline{2-2}
     &  \textbf{RQ4:} What are the challenges of ML data cleaning?\\ \cline{2-2}
     & \textbf{RQ5:} What are the challenges of data labelling faced by the ML application developers? \\ \cline{2-2}
    &  \textbf{RQ6:} What are the common approaches to validating data labelling by the ML practitioners? \\ \hline

\multirow{10}{*}{Feature Engineering} 
    & \textbf{RQ7:} How do ML practitioners identify class-imbalance in ML data and how do they ensure class-balance?\\ \cline{2-2}
    &  \textbf{RQ8:} What are the feature engineering techniques and tools commonly used by ML developers? \\ \cline{2-2}
    &  \textbf{RQ9:} What are the common limitations of the existing feature engineering tools and techniques?\\ \cline{2-2}
    & \textbf{RQ10:} What is the state-of-the-practice in feature quality assessment in ML application development?\\ \cline{2-2}
    & \textbf{RQ11:} What are the common practices for feature selection in ML application development? \\ \hline
\multirow{8}{*}{Model Building}	 
    & \textbf{RQ12:} What are the practices for ML model implementation commonly adopted by the practitioners? \\ \cline{2-2}
    & \textbf{RQ13:} What is the state-of-the-practice for ML model implementation testing by ML practitioners? \\ \cline{2-2}
    &  \textbf{RQ14:} What are the common symptoms that practitioners use to detect defects in an ML implementation? \\ \cline{2-2}
    &  \textbf{RQ15:} What are the practitioners perceived challenges of testing ML application? \\ \hline
\multirow{4}{*}{Model Management}
    & \textbf{RQ16:} What are the developers-perceived challenges of testing ML model deployment?\\ \cline{2-2}
    &  \textbf{RQ17:} What are the factors that ML developers commonly focus on during ML model management? \\ \hline
\end{tabular}
}
\end{table*}
\subsection{Survey Design}
To conduct the survey we defined an online questionnaire for the ML practitioners to participate anonymously. The questions in the questionnaire cover development activities of different phases of the ML application development life cycle. In addition, 
we asked the participants to report their technical skills, experience in ML and software development, job roles, and domains of their ML application development. 
The survey forms were made available to the interested participants through a web page. 
The survey has three parts as shown in Table \ref{tbl:Survey}. 
Part 1 collects some demographic 
information about the participants including the type of organization (e.g., industry or academia), job roles, skills, experience and ML domains of expertise. Part 2 of the questionnaire focuses on challenges and practices in the data collection, preprocessing and feature engineering. Part 3 of the questionnaire asks the participants about their development practices, tools, technologies and frameworks in ML model building, testing and deployment. 
All sections contain both open-ended and close-ended questions and also options to add comments 
by the participants where applicable. 
All the questions collectively meet the data requirements necessary to answer the research questions we defined in Table \ref{tab:Rq} for this study. 
In addition, an informed consent form was also available to the participants on the online survey page outlining the detail objectives, privacy and data use policy of the study. All queries and concerns of the potential participants were clarified by email responses from the authors. 

 \begin{table}
\centering
\caption{Structure of the Survey}  
\label{tbl:Survey}

\begin{tabular}{p{7.5cm}}
{\scriptsize
\begin{tcolorbox}

\begin{itemize}
    \item [1]\textbf{Part 1}
            \begin{itemize}
                \item [1.1]: \textbf{Organizational Information}
                    \begin{itemize}
                        \item [1.1.1]: Organization Type
                        \item [1.1.2]: Application Types
                        \item [1.1.3]: ML domain of specialization
                    \end{itemize}
                \item [1.2]: \textbf{Personal information and Experience}
                    \begin{itemize}
                        \item [1.2.1]: Software Development Experience
                        \item [1.2.2]: ML Experience
                        \item [1.2.3]: Educational qualification
                        \item [1.2.4]: Programming languages and frameworks
                        \item [1.2.5]: Roles and Responsibilities
                    \end{itemize}
            \end{itemize}
    \item [2] \textbf{Part 2}
            \begin{itemize}
                \item [2.1]: \textbf{Data for Machine Learning}
                    \begin{itemize}
                        \item [2.1.1]: Data Source
                        \item [2.1.2]: Data Quality- Attributes and evaluation
                        \item [2.1.3]: Data Cleaning- Approach and challenges
                        \item [2.1.4]: Data Labeling- tools and challenges
                    \end{itemize}
                \item [2.2]: \textbf{Feature Engineering (FE)}
                    \begin{itemize}
                        \item [2.2.1]: Feature Extraction- Tools,  Limitations
                        \item [2.2.2]: Challenges in FE for ML
                        \item [2.2.3]: Feature Selection- approaches
                        \item [2.2.4]: Feature Assessment/Validation
                    \end{itemize}
            \end{itemize}
    \item [3] \textbf{Part 3}
            \begin{itemize}
                \item [3.1] : \textbf{ML Model Training}
                    \begin{itemize}
                        \item [3.1.1]: Implementation Strategy
                        \item [3.1.2]: Testing Model Implementation
                        \item [3.1.3]: Model review and bug identification
                    \end{itemize}
                \item [3.2] : \textbf{ML Model Evaluation}
                \begin{itemize}
                    \item [3.2.1]: Testing Approaches and metrics
                    \item [3.2.2]: Challenges in Model testing
                \end{itemize}
                \item [3.3] : \textbf{ML Model Deployment}
                \begin{itemize}
                    \item [3.3.1]: Deployment Strategies
                    \item [3.3.2]: Integration and post-deployment testing
                \end{itemize}
                \item [3.4] : \textbf{ML Model Management/Monitoring}
                \begin{itemize}
                    \item [3.4.1]: Extent of Performance Monitoring
                    \item [3.4.2]: Performance Parameters
                    \item [3.4.2]: Performance Monitoring Challenges
                    \end{itemize}
            \end{itemize}
\end{itemize}
\end{tcolorbox}
}
\end{tabular}

\end{table}

\subsection{Data Collection}
To collect responses from the machine learning practitioners regarding our survey, we communicated the online link of the survey to the prospective participants by email along with our research objectives and requested their participation. Interested participants submitted their responses anonymously using the randomly generated participants' identification numbers. At the end of the survey deadline, we downloaded the responses of the participants. We used the participants' IDs in tracking and analyzing the anonymous survey data. 

\subsubsection{Selection of Participants}
We selected participants based on their self-declared profiles in the professional network LinkedIn. We also selected ML developers from the GitHub user community contributing to the development of ML applications. In both cases, we ensured that they are professionally attached to ML/AI application domains. For example, from LinkedIn, we selected users either based on their employment in different roles related to ML/AI application such as AI/ML engineer/developer, data scientist, AI/ML researcher/scientist, Software engineer, software architect, and PhD or Masters student in ML or relevant areas. For GitHub users, on the other hand, we select users from the list of contributors in ML/AI projects. 
In either case, our focus was to reach out to potential participants with expertise and experience in developing ML applications. Once selected, we requested the potential participants by email to participate in the online survey. We gave the necessary details on the objectives, procedures, and policies of the study and asked for their consent to participate voluntarily. 

We received responses from practitioners of diverse backgrounds. From about 700 requested potential participants, 81 respondents completed the survey which is about 11.57\%. To mention, out of the 81 respondents, all responded to \textit{Part 1} of the survey, 49 participants responded to \textit{Part 2} and 44 participants responded to \textit{Part 3} of the survey. We excluded responses of one participants with partial response to only \textit{Part 1} of the survey. So at the end, we retained the responses of 80 participants for our analysis.  



\subsection{Data Collection and Analysis}
Our survey was designed using Google forms and was made available to the respondents through a provided web link. We collected the data once the survey period was ended. 
We did some preprocessing of the responses to remove formatting or minor linguistic differences for correct analysis and descriptive statistics. 
To answer the research questions, we analyzed the data to compute descriptive statistics.
We then used visualization techniques to present the responses to have better insights into the trends, similarity, and contrast among different class of responses. 
For qualitative analysis of the responses from open-ended questions, we applied grounded theory \cite{Stol_2016, Charmaz_2006} based coding of the responses for categorization of the challenges and practices in different phases of the ML development. 
Here, we assigned qualitative coding for the segments of data from the participants' responses. This aims to make analytic interpretations of the concrete statements from the survey participants to compare, categorize, and summarize the responses. We named (coded) each distinct segment of data to develop abstract concepts for interpreting that data segment. The coding is to link data to an emerging theory that aims to explain the data. We started with initial coding that is open to possible concepts followed by more focused coding to organize or synthesize frequent initial codes. We did theoretical integration during focused coding and continue for subsequent steps to pinpoint the most salient categories from the data.    
Two of the authors performed classifications independently regarding the goals defined by the corresponding research questions. The authors resolved the disagreements observed in some cases by meeting in person to finalize the data classification. The classified data was further summarized based on analyzing the distributions and visualization. 
Based on the analysis, we summarized the practices and challenges in ML application development as reported by the survey participants.


\subsection{Privacy and Anonymity}
To ensure the privacy and anonymity of the participants, we did not collect any personal information. The participants were assigned a randomly generated code to use as the user ID. We use cookies to keep track of the returning user to assign the same user ID for different parts of the online survey. Participants were able to access the privacy and data usage policy along with the consent from for voluntary participation. Participants' data will be securely preserved for seven years. 
Participants were allowed to withdraw themselves and request data removal at any stage of their participation. 

\section{Results} \label{sec:results}

In this section, we present our results from the survey to answer the research questions. We also present our insights into the survey responses from the expert practitioners regarding the challenges and best practices in ML application development. We present our findings in the following subsections: 

\subsection{Demographic Distributions}
We summarize the demographic information of the participants as follows:
\subsubsection{Background}
\begin{figure}[h]
\center
\includegraphics[scale=0.65]{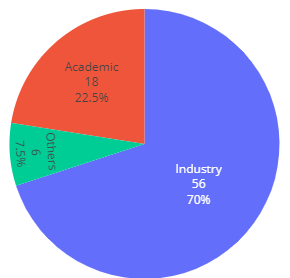}
\caption{Organization Types of the Participants}
\label{fig:Organization}
\end{figure}

\begin{figure}
\center
\includegraphics[scale=0.65]{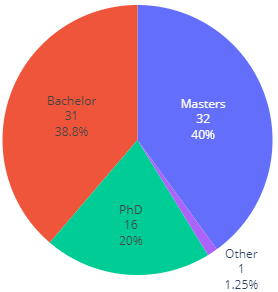}
\caption{Educational Qualifications of the Participants}
\label{fig:education}
\end{figure}

Among the 80 respondents who completed the survey, 56(70\%) participants are from the software industry, 18(22.5\%) from academia or research, 1(1.25\%) was with both academic and industry affiliation, and 5(6.25\%) participants identified themselves with other affiliations (Fig. \ref{fig:Organization}). The participants are from diverse academic background (Fig. \ref{fig:education}) comprising of 16 PhDs or above (20\%), 32 Masters (40\%), 31 Bachelors (38.75\%) and 1(1.25\%) mentioned as with ``Other" level of educational qualifications. 

\begin{figure}
\center
\includegraphics[scale=0.55]{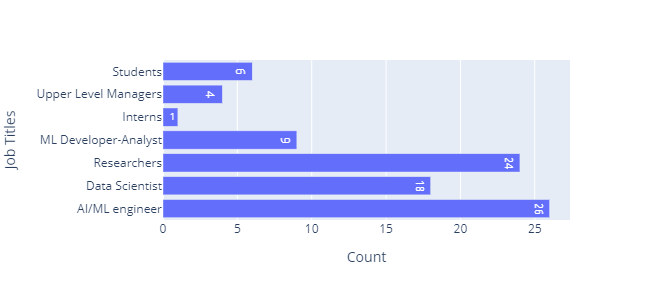}
\caption{Job Titles/Roles of the Participants}
\label{fig:jobtitles}
\end{figure}

The participants are from diverse roles (Fig. \ref{fig:jobtitles}) in their corresponding organization with 26(32.5\%) AI/ML engineer, 18(22.5\%) data scientist, 24(30\%) researcher with 10(12.5\%) of them identified themselves as AI/ML research scientist. Besides, 9(11.25\%) of the participants are with the roles of AI/ML developer/analyst, one (1.25\%) software development intern, and 4(5\%) with upper-level roles including one chief AI officer, ML software architect, software team lead, deep learning manager. In addition, the participants include three (3.75\%) PhD students, two (2.5\%) Masters students and one other student. 
The above diversity in the participants comprising both researchers and practitioners allows us to obtain a good representation of the skills and experience of varying levels.

\subsubsection{Professional experience}
\begin{figure}[h]
\center
\vspace{-0.2cm}
\includegraphics[scale=0.65]{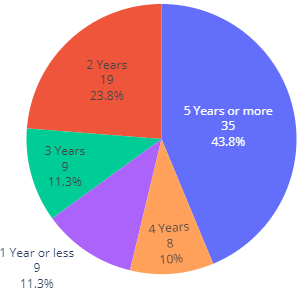}
\caption{Software Development Experience of the Participants}
\label{fig:SD-experience}
\vspace{-0.2cm}
\end{figure}

\begin{figure}[h]
\center
\includegraphics[scale=0.65]{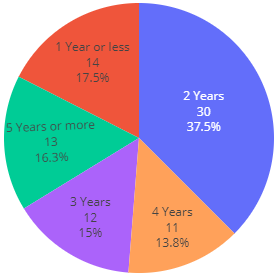}
\caption{ML Development Experience of the Participants}
\label{fig:ML-experience}
\end{figure}

As shown in Fig. \ref{fig:SD-experience}, the participants are highly experienced in software development with 53.8\% of them have a minimum 4 years of experience in software development. Among the participants, we have 35(43.8\%) participants who have worked for five years or more in software development and 8(10\%) with four years, 9(11.3\%) with three years, 19(23.8\%) with two years of experience respectively. Only 9(11.3\%) of the participants are relatively novice with less than 1 year of experience. The participants have diverse levels of experience in machine learning (Fig. \ref{fig:ML-experience}) with more than 80\% of the participants having at least two years of experience in machine learning application development. To be specific, 13(16.3\%) participants have five years or more experience in ML while 11(13.8\%) have four years, 12(15\%) have three years, 30(37.5\%) have two years, 14(17.5\%) are relatively novice with less than one year of experience in ML. 

It is important to note that there is a drop in the percentage of participants in higher experience categories. For example, participants with experience of five years or more dropped from 35(43.8\%) to 13(16.3\%) from software development to ML application development context. This could be explained as the migration of experienced developers from traditional software development to ML application development to adapt to the increasing AI/ML trends in the software industry. This is valuable to our study as such participants have wealth of knowledge and experience to compare and contrast the traditional software development and ML application development especially regarding the challenges and best practices. 
\subsubsection{Domains of expertise}
The survey participants work on developing applications in diverse machine learning domains. Our survey data shows that image processing and natural language processing (NLP) are the two domains with the top two number of participants, 45(56.25\%) and 44(55\%) from each respectively. Among the participants, 38(47.5\%) work in the area of predictive analytics and recommendation while 31(38.75\%) participants claimed to have working experience on clustering. Besides, 20(25\%) and 13(16.25\%)  participants work on video processing, and speech and audio processing respectively. Also, 3(3.75\%) of the participants use reinforcement learning (RL) in their ML applications while some other application domains of the participants include areas such as control and optimization, games, rendering and animation, security (anomaly detection), music generation, and biomedical engineering. Representation of participants from different application domains provides us with the opportunity to have developers' insights on the challenges and practices regarding the diverse area of machine learning and AI. 

\begin{figure}[h]
\vspace{-0.5cm}
\center
\includegraphics[scale=0.55]{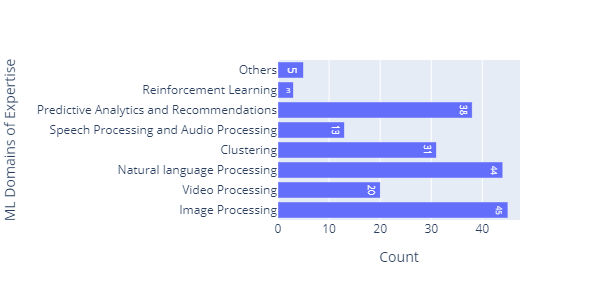}
\vspace{-0.5cm}
\caption{ML Domains of Expertise of the Practitioners}\label{fig:ml-domain}
\end{figure}

Participants have expertise in a diverse set of programming languages and technologies. Among the participants, 77(96.25\%) are Python users, which shows that Python is a remarkably popular language among ML practitioners. Besides Python, we have 16(20\%) C++ users,  11(13.75\%) R users, 10(12.5\%) Java users, 8(10\%) Matlab users and 6(7.5\%) SCALA users. 
In addition, a few participants claimed to use one or more of C\#, CUDA, STAN, JavaScript, Node JS, and Clojure as their languages in ML application development. 

\begin{figure}[h]
\center
\vspace{-0.25cm}
\includegraphics[scale=0.53]{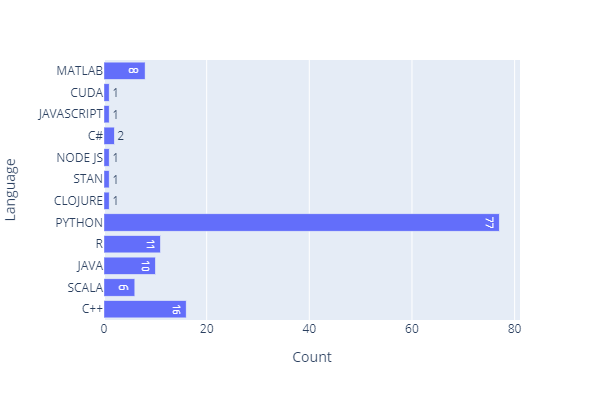}
\vspace{-0.85cm}
\caption{Programming Languages for ML Development}
\label{fig:language}
\end{figure}

\subsection{Trends in ML Application Development}
Here, we report the current trends in developing ML applications in the industry based on the response of the practitioners. We focus on the types of ML applications software industries are developing, software development methodologies, and the ML frameworks and tools developers are using to develop ML applications to answer the following research question:

\paragraph{\textbf{RQ1:} What are the current industry trends in developing ML applications?}

\subsubsection{ML Application Types}
\begin{figure}[h]
\vspace{-0.25cm}
\center
\includegraphics[scale=0.55]{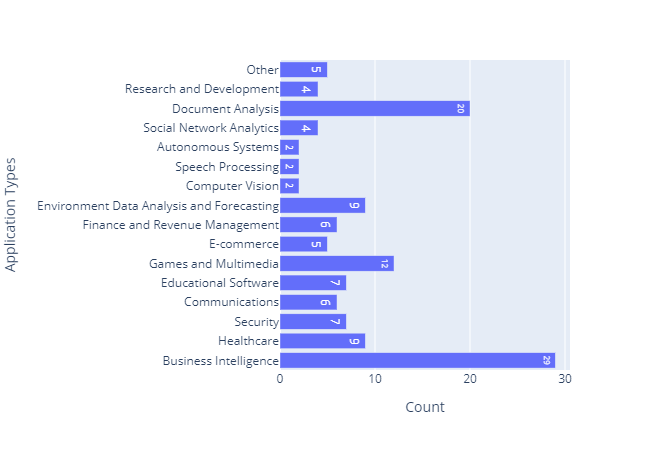}
\vspace{-0.85cm}
\caption{ML Application Types}\label{fig:application_type}
\end{figure}

Responses of the participants give an overview of the ongoing trend in the AI/ML industry regarding the types of applications developed (Fig. \ref{fig:application_type}). We asked the participants to list the types of AI applications commonly developed in their companies. We observe that companies are developing diverse classes of AI-based solutions encompassing different aspects of daily life, business, education, health, commutation, security, entertainment, research and innovation, social networking and so on. Based on the survey, we observe that software industries are highly focused on developing AI-based solutions for business intelligence (29 (36.25\%)). This is reasonable given the ongoing trends in the companies to leverage AI for improved products and services, customer clustering, product recommendations, prediction and forecasting for business decision support. The practitioners are also involved in document processing (20(25\%)) commonly based on the application of natural language processing. 
Companies are also developing solutions for entertainment (12(15\%)), healthcare (9(11.25\%)), education (7(8.75\%), security (7(8.75\%)) and communication (6(7.5\%)). 

Besides, there has been a considerable focus on developing ML-based solutions for business including E-commerce, finance, insurance, retails, and revenue management as 10 (12.5\%) of the participants reported these application types developed by their companies. Another important application area the practitioners are working on is environmental data analysis and forecasting as reported by 9(11.25\%) participants. Participants also reported working on building applications for social network analytics, control and automation such as self-driving cars and other areas of research and development in ML/AI including computer vision, speech processing, and simulation. So, our survey shows the diverse area ML/AI is being applied as the recent trends.  


\subsubsection{Software Development Methodologies}
As reported by the practitioners, agile software development methodologies have been widely adopted in software industries for ML application development. Among the participants 52(65\%) participants report that they use agile process for ML application development. Some widely used agile process frameworks used by the practitioners are namely SCRUM \cite{Schwaber_1997}, Kanban \cite{Anderson_2010}, and LEAN \cite{Poppendieck_2003}. Practitioners also reported the use of tools such as Jira\footnote{\url{www.atlassian.com/software/jira}} and Zenhub\footnote{\url{www.zenhub.com}} for the management of agile development process. Among the participating developers, 10 (12.5\%) reported to use other  data- or test-driven development process. A portion (18(22.5\%)) of participants reported that they do not use any specific development process for developing ML applications. 

\begin{figure}[h]
\center
\includegraphics[scale=0.5]{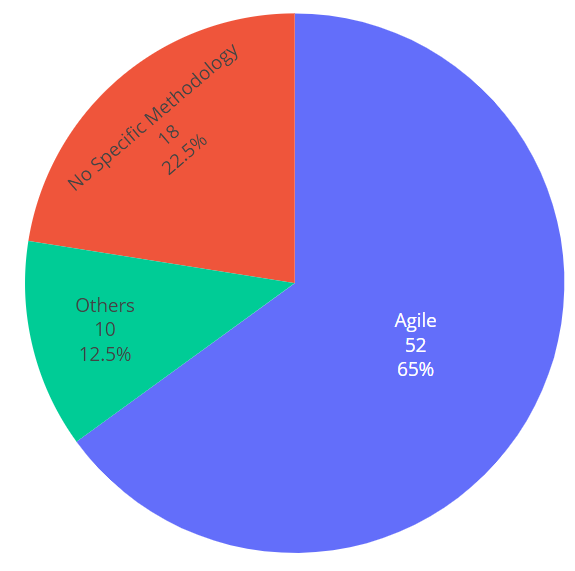}
\vspace{-0.25cm}
\caption{Software development methodologies used for ML Application Development}\label{fig:dev-methodology}
\vspace{-0.25cm}
\end{figure}

As mentioned by the practitioners, although agile process are the most commonly used, the development process is sometimes tailored to fit specific application development context, \ie ``\textit{agile/scrum but tailored towards ML model development processes}". Some practitioners refer to their agile development process as ``\textit{loosely organized agile}" or ``\textit{light agile}" and ``\textit{more explorative}". Depending on the context, some developers use either some adhoc or agile process for ML application development. They mentioned that ``\textit{many smaller-scale models are prototyped on an ad-hoc basis with no formal project methodology. Medium and larger projects borrow agile techniques}". While many practitioners do not use ``\textit{specific development process}", some prefer to use a data-driven or ``\textit{feature-driven}" or ``\textit{test-driven development}" development processes involving ``\textit{unit testing, integration testing, devops (continuous integration and delivery)}" for ML application development. Thus, we observe that practitioners mostly use agile methodologies for ML application development. However, the choice of development process may vary and the development process may require to be tailored to fit into specific ML application development needs.   
\vspace{-0.25cm}
\subsubsection{ML Frameworks and Tools}
\begin{figure}[h]
\center
\vspace{-0.8cm}
\includegraphics[scale=0.52]{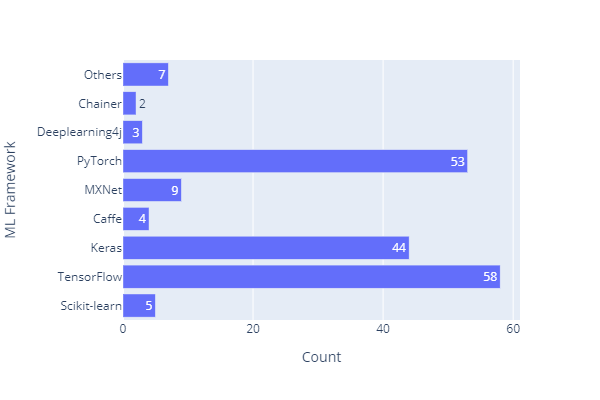}
\vspace{-0.8cm}
\caption{Commonly Used ML Frameworks for Application development}\label{fig:framework}
\end{figure}

From the responses of the participants, we have a list of popular ML frameworks and tools widely used by the ML practitioners (Fig. \ref{fig:framework}). Among the respondents, 58(72.5\%) use TensorFlow as their ML framework for application development showing it as the most popular framework in AI/ML application development. Then, 53(66.25\%) of the participants reported that they use PyTorch making it the second-highest popular ML framework followed by Keras, a high-level ML framework based on TensorFlow which is reported to be used by 44(55\%) participants. Among other ML frameworks MXNet, Scikit-learn, Caffe, and Deeplearning4j are reported to be used by 9(11.25\%), 5(6.25\%), 4(5\%) and 3(3.75\%) participants respectively. Some participants have also reported that they use frameworks like Chainer, Tensorflow.js, Caret, OpenCV, ML.Net, XGBoost, MLlib for their ML application development. It is to be noted that each participant may use multiple frameworks for ML application development and thus the count of participants for different frameworks are not mutually exclusive. 


\begin{tcolorbox}
\textit{\textbf{Finding 1}}: As reported by the practitioners, the trends in ML application development show that (1) Business Intelligence (BI) systems is at the key focus of ML application development. Other ML application types include (but are not limited to) healthcare, security, document processing, entertainment covering wide areas for human life and business, (2) practitioners widely use agile software development methodologies to develop ML applications, and  (3) TensorFlow is the most widely used ML framework followed by PyTorch and Keras. However, ML developers use different development processes and frameworks based on their specific ML development context.   
\end{tcolorbox}
\vspace{-0.5cm}

\subsection{ML Data Collection and Pre-processing}
Machine Learning applications are data-driven, and so it is intuitive that the quality of the input data is very important for the performance of the ML models. Based on the responses from our survey participants we compile different processing tasks, common practices in ML data preparation. From the responses, we know the state of practices adopted by the ML practitioners. We summarize the common practices and challenges related to ML data processing as in the following:

\begin{figure}[h]
\vspace{-0.75cm}
\center
\includegraphics[scale=0.5]{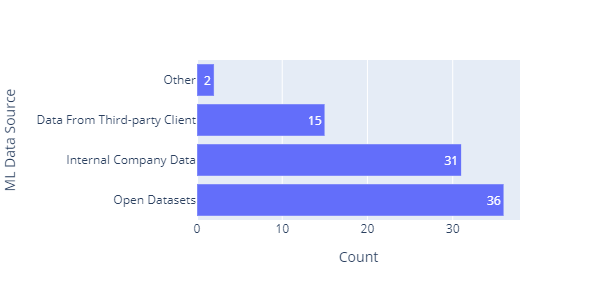}
\vspace{-0.85cm}
\caption{ML Data Sources}\label{fig:data-source}
\end{figure}

\subsubsection{ML Data Sources} 
Depending on the ML application domains, the types of the data may vary widely as well as the sources of the data. Data can be of different forms such as text, images, videos, speech, business transactions, time-series data and so on. Similarly, these data may come from different private or publicly available sources (Fig. \ref{fig:data-source}). As mentioned by the survey participants, companies rely on one or more sources for ML datasets for their ML application development. One of the common sources of ML data is the open-source data sets made publicly available by different academic institutions, companies, various tech and research communities (\eg~ Arxiv, Kaggle). 
As mentioned by the participants, companies rely on internal company data for developing ML solutions either for themselves or for others. Many software companies develop custom ML solutions for their third party clients based on their supplied data regarding business transactions, users, and the data collected from internal operations or even external environments using sensors over a certain period of time. ML data is also collected from online sources by web crawling and scraping. 

 To summarize, Open-source data sets are the leading source of data for ML application development. Besides, private data and data from third-party clients are also common sources of ML data as reported by the practitioners. 
 



\subsubsection{RQ2: In practitioner's perception, what are the important quality attributes of ML data?}

ML models are data-driven and so the quality of the data is important for the performance of the ML models and consequently the applications containing the ML models. We asked the practitioners about this important topic to learn about the quality attributes that ML developers focus on in practice while assessing data quality. We then compile and classify the data quality attributes based on the responses of the survey participants. We list the key observed quality requirements of the ML data as pointed out by the practitioners as follows:

\paragraph{Feature representativeness:} 
In machine learning, the primary purpose of the data is to train the ML models. For this, the data must be representative of the necessary discriminative features to learn from. Thus, how well data represent the characteristics capable of differentiating different hidden patterns in the data is very important. Practitioners thus emphasize on ``\textit{feature quality}" which requires ``\textit{high discrimination between features}". This can be assessed by statistical measurements on the data set such as ``\textit{balanced distribution}", ``\textit{high variance}, and ``\textit{low correlation}" among the features and with the ``\textit{target}" variable(s).

\paragraph{Adequacy:}
ML models need an adequate amount of data samples for training. In practitioners' word ML models need ``\textit{lots of samples with wide variation, equal(ly) distributed across fields{/}classes}". The adequacy of the data is hard to define and depends on different factors such as the data, problem, number of features, number of distinct classes, and ML algorithms. 

\paragraph{Diversity:}
ML models need to have ``diversity'' regarding 
the coverage and distribution of data among different classes present in the data set. The practitioners have emphasized on the diversity mentioning that the data should contain ``\textit{...samples with wide variation, equal(ly) distributed across fields{/}classes.}". The practitioners also emphasized on the ``\textit{distribution of response variables, (and) distribution of each features}". They also mentioned  ``\textit{subject area coverage, sampling uniformity, sparsity, vocabulary coverage}" as important characteristics that enhance the diversity in the data set. Like other data quality characteristics, different diversity factors and their importance may vary with data, problem and the algorithms. 

\paragraph{Labelling Accuracy:}
Labelling accuracy is very important for the ML data set. So, it is important to ensure that there are ``\textit{no mislabelled data. The dataset should be treated with the utmost care, because a bad dataset means a bad model even if it’s trained well.}" Practitioners thus emphasize on data ``\textit{quantity and correct labels}". ``\textit{The quality of the labels} \ie ``\textit{reliability of (data) annotations}" is very important and ``\textit{the structure, accuracy and quality of information would play a large role in determining the importance of the ML data sets}." So, in practitioners' word ``\textit{(labelling) consistency is very important; for a particular field I was working on a year ago, there were only two available data sets, but both of them had inconsistent labeling, which made them unusable.}"

\paragraph{Completeness:}
Machine learning data need to be complete meaning that there should not be missing values in the data or at least there should be ``\textit{enough data with minimum missing values}". Data samples with missing values are either dropped or some transformations are applied to fill in the missing values with the best approximate values. 

\paragraph{Consistency:} 
Like the adequacy of the ML data, it is very important for the data to be consistent. The consistency of the data can be in terms of the correctness of values, data types or the format or structure of the data or even the labelling. 
The practitioners thus focus on the ``\textit{structure, accuracy and quality of information}". ML data need to be ``\textit{consistent with inference data; be relevant for the model; be consistent with itself.}". Consistency defines the suitability of the data to use in the ML models. 

\paragraph{Reliability:} 
ML data need to be reliable, meaning not only correctness and consistency but also the reliability of data source, data collection and annotation procedure. The reliability of ML data can be validated by different cross-validation processes. The practitioners suggest checking if ``\textit{it (the data) has been verified by multiple sources}". It is crucial especially in health and safety critical domain such as for ``\textit{medical data}". ML data to be reliable, practitioners expect that the ``\textit{data is clean, well explained, come from good annotations. You (developers) also need to know how the data was generated.}".

\paragraph{Noise Level:} 
ML data can have noise in it due to missing or erroneous values and outliers in the data. The data can be incorrect in terms of values or data types. Thus, ML data require different transformation and cleaning to remove noises and to improve data quality. 

\paragraph{Relevance:}
ML data need to be relevant for the problem, \ie the data should represent the necessary characteristics meaning the  ``\textit{existence of viable features}"  that ML models can learn from. Like other data quality requirements, the relevance of the data ``\textit{depends on the problem}". 

\paragraph{Class Balance:}
For ML data, \textit{class balance} is crucial for the accuracy of the ML model. For an unbalanced data set, the model is likely to be biased to the majority class, leading to poor accuracy, especially for the minority class. Practitioners recommend the data ``\textit{samples (to be) well balanced across classes}" \ie~data is ``\textit{equal(ly) distributed across fields/classes}".

\paragraph{Distribution:} 
ML data should have balanced distribution across the classes and have ``\textit{sampling uniformity}". Different statistical measures (\ie descriptive statistics), variance, correlation are commonly used by the practitioners to measure data relation and distribution. 

\paragraph{Performance impact:}
One of the key concerns is how well the model performs based on the given training data. The quality of the data is thus also reflected in the performance of the model. For such quality assessment, practitioners often build a prototype model based on the subset of data and measure model performance such as ``\textit{AUC ROC on test set}". 

\paragraph{Low Bias:}  
There can be different sources of biases in the ML data set. The biases can originate from the human error or perception differences of it can be from the data historically containing discrimination or biases in it. The biases should be eliminated from the data set as much as possible. Thus, the practitioners recommend that ML data need to be ``\textit{diverse, not biased}". 

\vspace{0.2cm}
\begin{tcolorbox}
\textit{\textbf{Finding 2}}: Based on the practitioners' responses, the key characteristics of ML data are: feature representation, adequacy, diversity, labelling accuracy, completeness, consistency, reliability, noise level, relevance, class balance, data distribution, performance impact, and biasness. ML data should maintain these quality attributes to build high-performance models.    
\end{tcolorbox}

\subsubsection{RQ3: What is the state-of-the-practice regarding the data processing tasks, techniques and tools for quality assurance of ML data?}
Quality of the dataset is one of the key factors that contribute to the performance of the ML models. Here, we discuss common data processing tasks,  techniques, and tools for ML data processing for quality assurance of ML data. 

\paragraph{Data Processing Tasks:}
Practitioners may need to employ a series of preprocessing and transformation to ensure the desired quality of the data or ML models in turn. Based on the practices reported by our survey respondents, we can broadly group the data processing task into the following:

\begin{itemize}
    \item \textit{Data Transformation}: ML practitioners often need to apply different transformations on the dataset to prepare for machine learning algorithms. These transformations may include simple corrective transformation such as adjusting the data types or structure of the data. Data may also need some advanced transformations like reducing the dimensions of the data while preserving its key characteristics or hidden patterns. ML data often require normalization and scaling to transform the values to a range suitable for ML algorithms. Another important quality attribute of ML data is the class balance, which can affect model performance. In such a case, some practitioners reported that different boosting and re-sampling techniques are used to remove class imbalance problems in the ML dataset. 

    \item \textit{Data Analysis}: To analyze and assure the quality of ML data, practitioners employ different analysis techniques. The first step in quality assurance is to understand the dataset regarding the distribution and basic trends. Practitioners commonly do a manual analysis to have the basic perception of the data characteristics. Another common approach as mentioned by the practitioners is the visualization of the data. The common visualization techniques include the presentation of data using different charts and graphs. Some practitioners also use advanced visualization techniques such as t-SNE \cite{Maaten_2008} that facilitates the visualization of multidimensional data in a more flexible and elegant way. In our survey, practitioners also reported that they use exploratory data analysis to evaluate data quality. This analysis helps to understand the common characteristics, category, and trends in the dataset. Another common approach to data quality assurance is to perform statistical analysis or to cluster data to understand the distributions and trends in the ML data set. The analysis can be performed on randomly selected samples from the data set or on the whole dataset. 
    Another approach to assess ML data quality is to build a prototype model based on a subset of the data and verify the model performance. 
    The type and extent of analysis may depend on the problem, data and specific objectives of the data analysis. 
\end{itemize}

\paragraph{Tools and Techniques for ML Data processing:}

The practitioners depend on different tools and techniques for ML data analysis. One common technique reported by our survey participants is the manual inspection of the data. Manual inspection is a reliable technique as the developers can take advantage of their domain knowledge to assess the quality of the ML data to perceive the common patterns in the dataset. ML data may also need to be annotated manually for categorization and labelling. However, manual analysis is likely to be costly and may suffer from scalability issues in case of a large data set. Another approach commonly used by the practitioners is to visualize the data set. As reported by the survey participants, open-source tool \textit{Jupyter Notebook}\footnote{https://jupyter.org/} is a widely used tool for data exploration and visualization. Practitioners also reported using other commercial data analysis tools (\eg~\textit{Kibana}\footnote{https://www.elastic.co/kibana}) for exploratory data analysis and visualization. Practitioners also reported that they use \textit{Apache Spark}\footnote{https://spark.apache.org/} for ML data processing especially in the big data context. 

Another technique used by the developers is to write custom scripts for data analysis and visualization using descriptive statistics, charts, and graphs. Custom scripts can also be used to fix for missing and duplicate values, to identify data types and value range inconsistencies, detection of labelling errors, and for checking data structures or formats. One important point to note is the fact that many practitioners reported that they do not use specific tools for data quality analysis and some times do not even check data quality, and instead rely on assumed quality based on the source of the data. However, this reliance may fail to identify potential issues in ML data quality and may consequently lead to poor quality ML models. However, despite of different commonly used tools and techniques, domain knowledge plays an important role in the application of tools and techniques and the effectiveness of ML data quality assurance. 

\paragraph{Common Practices in ML Data Processing:}

One of the key challenges of the ML data collection and preprocessing is that the data and the necessary processing can be domain and problem-specific. Thus, no specific tool may fit all the problems or data processing requirements. The responses from our survey participants also reflect the challenges of dealing with these variabilities. Overall 76.6\% of the participants mentioned that they do not use a very specific tool for ML data analysis. One of the key reasons is likely to be the above-mentioned fact that one specific tool is not capable of handling diverse data analysis requirements and practitioners may use very domain or problem specific tools and techniques. It can also be explained by the limited availability of data analysis tools with comprehensive features to cover the processing of data from diverse domains as only 14.9\% of the practitioners have reported to using specific data analysis tools. Besides, some of the participants reported that they rely on existing Python libraries and frameworks to develop their custom data analysis scripts. Thus, it is important to develop necessary tools for data analysis with comprehensive coverage of data analysis requirements in diverse problem settings. 

\begin{tcolorbox}
\textit{\textbf{Finding 3}}: For quality assurance of ML data, practitioners apply different data transformation operations such as noise removal, replacement of missing values, dimensionality reduction, class-balancing and normalization. ML practitioners also use different tools and techniques for analysis and visualization of data for better understanding. However, more than two-third (76.6\%) of the developers do not use specific tools rather use diverse data or problem specific techniques or develop their own customized solutions.     
\end{tcolorbox}


\subsubsection{RQ4: What are the challenges of ML data cleaning?}
Cleaning ML data is an important data preprocessing step to remove noise from the ML dataset. Based on the practitioners' responses we list the following challenges in ML data cleaning: 
\paragraph{Generalization:} Data cleaning like most other tasks in the ML application development workflow is hard to generalize as it is usually ``\textit{geared towards specific applications}". This is due to the inherent domain- and problem-specific variations in data, ML frameworks, and algorithms, and even the target application platforms. This has also been reflected in the practitioners' responses as one respondent mentioned ``\textit{There is no one-size-fits-all tool and probably will never be one.}". Another respondent mentioned ``\textit{...it is practically impossible to make a general tool, as it depends on the data and the problem at hand.}". So, ``\textit{they are not generalizable to different use cases, like text and images.}". One common practice adopted by the ML developers is to develop or customize their data cleaning solutions as mentioned by one respondent: ``\textit{Sometimes they are not adaptive enough for my problems so I have to write my own.}" 

\paragraph{Scalability:} Another key challenge in data cleaning as reported by the survey practitioners is the ``\textit{scalability to big data sets.}". Most tools and techniques may suffer from the \textit{scalability} issues. This challenge is intuitively understandable particularly because of the rapidly growing volume of ML data. The data volume may often exceed the processing memory (``\textit{scaling to multi terabytes.}"). Thus, practitioners need to devise custom techniques to process larger data set in small capacity machines under resource constraints. Otherwise, it may impact the data processing cost due to large data processing resource requirements. 

\paragraph{Automation:}  
Some practitioners feel the need for ``\textit{automated analysis}" for data cleaning and reported that current data cleaning techniques are ``\textit{poorly automated}". However, practitioners are aware that ``\textit{some tasks cannot be automated...}" 
and recommend that ``\textit{...rule-based and AI/ML techniques need to be applied to data cleaning itself}". 
This suggests the idea that ML techniques can potentially be applied to automate the data cleaning tasks. Data regarding the cleaning techniques applied to existing ML applications are likely to be leveraged.    
Due to various diversities in data and problems, it is challenging to integrate the data cleaning and processing tasks into the ML workflow which further limits the automation of the data cleaning and other preprocessing tasks. 

\paragraph{Data quality:}
``\textit{Most data is noise (noisy)}" and thus cleaning of these types of data can be costly. Moreover, data can be from different sources and in different forms and so their levels of quality. For example, text data can be with different encoding schemes while image data can be in different formats and quality. 
When the data is too noisy, the cleaning task becomes costlier and often impossible given the tools and techniques available. 
Data from companies are proprietary data and the structure of the data is likely to be driven by other business or technical factors than the application of ML.  

\paragraph{Lack of standard:}
Another issue the practitioners commonly face is that there is no defined standard of ``clean-data". The cleanliness can be relative and may vary with data, problems, and algorithms. This makes it harder to devise robust techniques for data cleaning. 


\paragraph{Efforts and Costs:}
Data cleaning can be costly (``\textit{It’s very time and labour intensive}") and thus ``\textit{requires a lot of efforts}", time and computational resource requirements. Also, data processing tasks can be highly iterative and the continuous expansion of the data may trigger repetitive data processing incurring high cost. 

\paragraph{Lack of Tools and Features:}
As the data types and the required processing may differ widely from one problem to another, tools are likely to be with a domain or problem specific features. This limits the adaptability of tools for diverse ML data. The lack of features and data dependencies limit the usability of the data cleaning tools and techniques. The practitioners also mentioned ``\textit{the high complexity of use}" as a challenge to using data cleaning tools effectively. Also, ``\textit{sometimes they (tools) are not adaptive enough for my (specific) problems...}" and this lack of flexibility also limits the use of tools for processing ML data of diverse characteristics.

\paragraph{Context and perception differences:}
From the responses of 
our survey participants, we observe a difference of perception on the challenges of data cleaning between the ML practitioners. 
This difference of perception is likely due to the different contexts in which they performed 
their data cleaning tasks. 
The response ``\textit{Till now, from my usage experience, I didn't find any limitations on the tools I have used during my projects.}" is thus likely to represent a domain and context-specific view of the respondent and may or may not apply to the development contexts of other participating practitioners. 


\paragraph {Requirement for domain expertise:} ML data processing requires a clear understanding of the data, target problem, and algorithms. However, understanding the structure and semantics of the data from a particular domain may often require basic and sometimes advanced knowledge in the domain. For example, to process natural language texts in the mental health domain the ML practitioners are in ``\textit{... need for expertise in linguistics and mental healthcare.}" The requirement for domain expertise may vary across the domains and the type of the problem being addressed by the ML application. 

\vspace{0.2cm}
\begin{tcolorbox}
\textit{\textbf{Finding 4}}: From practitioners' view, the key challenges in ML data cleaning approaches are: generalization, scalability, automation, data quality, lack of standard, required efforts and costs, lack of tools and tool-features, domain knowledge requirement. From practitioners' responses, it is observed that ML developers are sometimes not aware of the importance or domain-specific challenges in data cleaning.  
\end{tcolorbox}





\subsubsection{RQ5: What are the challenges of data labelling faced by the ML application developers?}
Feature labelling is very important as incorrect labelling affects model accuracy. However, labelling of features is a challenging task especially when data volume is large and due to different domain- and problem- specific requirements and constraints. Based on the responses from our survey participants, we identify several key challenges perceived by the ML practitioners as follows:

\paragraph{Data Volume:} 
One of the key challenges in feature labelling reported by the practitioners is the large volume of data. Labelling commonly involves manual effort and given the increasing volume of ML data, labelling can be very challenging due to the constraint ``\textit{large dataset vs limited human resources}". Because of the data volume, ``\textit{the amount of work to be done can be overwhelming}". 

\paragraph{Cost:}
 ``\textit{Labeling the data is quite hectic and time taking process}" and thus may incur ``\textit{high cash and labor costs}". ``\textit{Also in some cases, labeling the data also requires a lot of knowledge of the field to which the data belongs}". However, ``\textit{using experts for labeling is expensive}" while ``\textit{using non-expert for labeling results in low quality training set}". Again, ``\textit{it is better to have opinions of several experts rather than a singleton labeling to avoid biased opinions and ensure (the) validity of the labeling}". This further increases the cost of the labelling of the ML dataset. 
 
\paragraph{Required Domain Expertise:}
Data labelling often requires expert-level domain knowledge. For example, labelling data in medical imaging such as diabetic retinopathy requires ``\textit{a super-skilled workforce, such as doctors to estimate the level of diabetic retinopathy from images}". However, ``\textit{domain experts are hard to reach}". So, the requirements for domain expertise in ML data labelling not only make the labelling task challenging but also make labelling excessively expensive. 
However, the required level of domain expertise may vary across domains. 

\paragraph{Automation:}
Labelling of ML data is mostly manual or ``\textit{poorly automated}". ``\textit{Manual labeling can be very frustrating and time taking}" and thus ``\textit{labeling is slow and expensive}". 
For automation, standard procedure is necessary. However, due to diverse variabilities involved, ``\textit{coming up with a standard annotation procedure}" is quite challenging. 
``\textit{It is very hard to make a criteria that can be validated by scripts or other automated tools}". Thus, although ``\textit{human annotation is expensive}" there is a ``\textit{lack of tools for annotation}". Also, for automation, it is necessary to have ground truth for validation. However, there is also ``\textit{lack of clear ground truth}". 

\paragraph{Domain Dependency:} 
ML Data and the labelling objective may vary widely across different application domains. As the practitioners claimed, data labelling ``\textit{becomes more difficult is(as) the dataset is domain specific}". Thus, the data labelling criteria and the required expert-level knowledge is also very domain specific. For example, ``\textit{for legal documents, lawyers are best suited to annotate}". This domain dependency puts a limit on the human labelling experience to be transferable to other domains. 

\paragraph{Biases:}  
One of the key challenges in ML data labelling is the potential biases or inconsistencies. There are different sources of possible biases or errors in data labelling. As data labelling is manual in most cases, ``\textit{discrepancies among humans}" \ie~the differences in knowledge and perception among labellers can introduce labelling biases or inconsistencies due to ``\textit{subjectivity}" as ``\textit{everyone has their own point of view}". Again, the annotators may ``\textit{have no understanding of the importance of the quality or lack proper training, so the labeling is inconsistent}". 

\paragraph{Data Quality:}
The quality of the data also has impact on the data labelling. Too much noises in the data and incompleteness of data due to missing values can affect the labelling. There can be multiple labels for single data and to avoid label confusion requires clear labelling guidelines. 

\paragraph{Reliability:}
Assuring the reliability of the ML data labelling is another challenging task. Due to the large volume of data, labelling is likely to require team-efforts. Perception difference among the team members may result in inconsistent labelling. 
Again, due to the overwhelming volume of the task ``\textit{ML researchers often rely on third-party annotations}". However, as mentioned ``\textit{third party annotators have no understanding of the importance of the quality or lack proper training, so the labeling is inconsistent}". 
Again, the practitioners claimed that often ``\textit{workers are not trust-worthy}", so it is challenging to make feature labelling reliable. Practitioners recommend cross-validation by multiple labellers as a remedy for the reliability risk of feature labelling. 

\paragraph{Lack of Guidelines:}
One of the important challenges the practitioners mentioned is that there is no comprehensive guidelines on the feature labelling. Practitioners expressed the ``\textit{..need to have a strict guideline for labelling}". However, ``\textit{it is very hard to make a criteria that can be validated by scripts}" given the variabilities involved in specific ML domain. 


\begin{tcolorbox}
\textit{\textbf{Finding 5}}: From practitioners' view, the key challenges in data or feature labelling are: large data volume, cost, domain expertise requirements, automation, domain dependency, biases, data quality and ensuring labeling reliability. Practitioners also expressed their need for comprehensive guidelines for feature labeling. 
\end{tcolorbox}

\subsubsection{RQ6. What are the common approaches to validating data labelling by the ML practitioners?}

Accuracy of ML data labelling is very important as incorrect labelling can have adverse effects on ML model performance. It is thus important to know the current practice of the ML developers for testing the data labelling accuracy. Based on the responses of our survey participants, we see some common practices used by the ML practitioners. As there is a lack of automated tools for labelling validation and it may require manual checking by domain experts, \textit{Manual Investigation} is mentioned by most practitioners (35(71.42\%)) as the approach for label validation. Besides, 22 (44.9\%) practitioners reported that they use automated tools or scripts for testing data labelling accuracy. However, in many ML domains, the data labelling is not required rather the neural networks is capable to learn from the data. Some practitioners build a prototype model on the subset of data and evaluate the data labelling accuracy based on the performance of the prototype model. In both automated and manual label validation, domain knowledge plays an important role. 

\vspace{0.2cm}
\begin{tcolorbox}
\textit{\textbf{Finding 6}}: Manual investigation is reported to be a common approach for validating feature labelling while some practitioners use tools and automated scripts. Domain knowledge is important for automated and manual validation of feature labels.  
\end{tcolorbox}

\subsection{Feature Engineering}

Feature engineering is one of the key steps in ML workflow unless the feature is learned automatically from the data such as in deep learning. In our survey, we asked the practitioners about how they do feature engineering to learn about the current state of practice. We report on the tools, techniques and challenges of feature engineering in the following subsections:

\subsubsection{RQ7: How do ML practitioners identify class-imbalance in ML data and how do they ensure class-balance?}
In machine learning especially in supervised learning context \textit{class imbalance} is a crucial issue as it can severely affect the performance of the ML models. Thus, it is important to identify the class imbalance and to take the necessary approach for balancing the ML data set regarding different representative classes. 
The first step to resolve the class imbalance of the ML data set is to identify the balancing issue in the first place. ML practitioners have shared different practices they use to assess and identify class imbalance in the data set. 
\paragraph{Identification of class-imbalance:}
We list the common approaches for identifying class imbalance as follows: 

\begin{itemize}
\item Statistical Analysis: 
One of the commonly used practices to identify class balance in the ML data set is to statistically analyze the data set. Descriptive statistics and the class distribution can indicate how well data is balanced across different classes.  

\item Data Visualization: 
Visualizing data is another approach widely used by the ML practitioners for testing class balance in the data set. In practitioners' word, ``\textit{testing (class balance) can be done by statistical approach or by visualizing the data on a graph which will show how balanced the data is}".  

\item Sampling and Analyzing Data Subset: 
Another approach to testing the class balance is to analyze randomly sampled subset of data as mentioned by the practitioners ``\textit{(we) randomly sample a portion and calculate the class distribution in it}". 

\item Manual Verification: 
Practitioners reported to use manual analysis to check the balance or distribution of the ML data. However, manually checking data can be time-consuming, costly and may not be scalable when data volume is large.  

\item Model Performance: 
Some practitioners mentioned a kind of reactive approach to testing class balance. Here, instead of balancing data proactively before training, their approach is to observe the impact of the imbalance on the model first then balance data if necessary. As mentioned by one of the practitioners where s/he first builds ``\textit{baseline models like linear regression and logistic regression to see how good/bad the imbalance affects the predictions}". Similarly, another practitioner mentioned the way as to ``\textit{evaluate sensitivity of the model at the end on generated data}". Some practitioners also reported measuring model performance based on k-fold cross-validation to assess the impact of class imbalance.   
\end{itemize}

In some ML domains, no explicit data labelling is necessary or the data is naturally imbalanced depending on the domain and thus the class imbalance issue may not apply to those contexts, and it has also been mentioned by different practitioners. 


\paragraph{Techniques for class-balancing: }
Based on the responses from the practitioners, we list some commonly used techniques for balancing data set as follows: 

\begin{itemize}
\item Data Re-sampling: 
The resampling of data is a commonly used technique to balance the ML dataset. To balance data, either minority class can be up-sampled or the majority class can be down-sampled. The balanced data then can be used for ML models. Practitioners reported the use of tools like SMOTE \cite{Chawla_2002} or ADASYN \cite{He_2008}.  

\item Stratification: 
Practitioners also reported using stratification techniques to understand and balance the data. A practitioner stated that: ``\textit{we write our own stratification solutions, based on the descriptive analysis}". Practitioners also mentioned applying stratified split of training and test data to reduce the impacts of class imbalance. Also, one of the practitioners mentioned data augmentation, stating that ``\textit{...will utilize data augmentation methods if class balancing does not satisfy requirements}" without specifying specific augmentation technique. 
\end{itemize}
The requirements and the techniques for balancing data are likely to depend on the particular ML context regarding the problem domain, the data and the ML algorithms. 

\begin{tcolorbox}
\textit{\textbf{Finding 7}}: Practitioners reported to use  statistical analysis, data visualization, analyzing randomly sampled data, manual verification, and measuring model performance to test class balancing of labelled data. To ensure class-balance, practitioners commonly perform data re-sampling (up or down sampling) and stratification of distribution for class balancing.    
\end{tcolorbox}

\subsubsection{RQ8: What are the feature engineering techniques and tools commonly used by ML developers?}

Based on the responses from the participants, we identify the following techniques and tools for feature engineering commonly used by the ML practitioners: 

\paragraph{Manual analysis:}
Based on the data and the ML problem domain, features may need to be ``\textit{hand crafted}". As one practitioner mentioned ``\textit{I need to learn about how the data is generated and formulate the features that will be useful}". 

\paragraph{Custom Programming:}
Feature for ML models are likely to be problem dependant, thus ML practitioners decide on features based on their domain knowledge and devise custom techniques for feature extraction as reflected in the response ``\textit{Domain knowledge remains my favorite tool, I understand the problem and read the current research to devise the best features}". Also, user-written custom scripts in addition to existing libraries and tools can be used for feature extraction as mentioned by practitioners;  ``\textit{we use custom scripts and functions in addition to NLP libraries like nltk, spacy and tensorflow}".

\paragraph{Using Libraries and Frameworks:}
One widely used practice among the ML developers is to develop feature extraction functionalities based on the available libraries and frameworks. Based on the responses of the survey participants, \textit{scikit-learn} is a very widely used ML library for feature extraction. Besides, some other frequently used Python libraries are \textit{pandas, numpy, scipy, and networkx} while for natural language processing (NLP) tasks \textit{nltk} and \textit{spacy} are among the widely used python libraries. R and Matlab based libraries are also used by the practitioners for ML feature extraction. For computer vision domain practitioners reported to using CNN or OpenCV for image feature extraction. The practitioners also reported that they use frameworks like TensorFlow and PyTorch.

\paragraph{Using Feature Engineering Tools:}
Some practitioners have reported the use of available tools for feature extraction. For example, one respondent mentioned the use of \textit{Data-Miner}\footnote{https://data-miner.io/} and then \textit{Featuretools}\footnote{https://www.featuretools.com/} for feature extraction. Some other tools as reported by the participating practitioners are DataVec, Weka, XGBoost for feature importance, PCA and Fourier transformation for dimension reduction. However, the use and suitability of tools vary with the problem domain and the data.

\paragraph{No Feature Engineering:}
Practitioners using deep learning based techniques do not require specific feature extraction technique as the network is expected to learn from the data.  

\vspace{0.2cm}
\begin{tcolorbox}
\textit{\textbf{Finding 8}}: Based on the practitioners' perception, the commonly used techniques for feature extraction are manual analysis, custom programming using existing libraries and frameworks for data processing, and Using available feature engineering tools. However, deep learning techniques may learn from the data without explicit feature engineering.  
\end{tcolorbox}

\subsubsection{RQ9: What are the common limitations of the existing feature engineering tools and techniques?}

Although tools and methods for feature extraction are very useful for ML practitioners, there are a limited number of tools available. Existing tools also do not cover all diverse requirements in feature engineering. The participating ML practitioners have mentioned different limitations of the existing feature engineering tools and techniques as follows:

\paragraph{Generalization:}
\textit{Generalization} is one of the key limitations of the existing feature extraction methods and tools since ``\textit{every problem is different}". And ``\textit{for most applications, the shape of the data that has to be input to the model is highly specific to the internal company data that we cannot use out of box tools to help us easily automate feature engineering}". So, due to the inherent variabilities in ML problems, data and algorithms, it is very challenging for any feature engineering tool to generalize for diverse problem contexts. 

\paragraph{Scalability:}
Another limitation of the feature extraction tools pointed out by the survey participants is that ``\textit{they (tools) are not scalable}" and thus ``\textit{hard to visualize quickly and efficiently on large data sets}". They also mentioned that 
``\textit{the current feature engineering tools is that they are not well parallelized, and when other parallelization libraries are introduced like multi-processing many problems arise causing me to edit the library and fundamentally having to change the code.}". This affects the performance and limits the application of tools especially for processing larger-scale data. 

\paragraph{Automation:}
Practitioners also mentioned that 
feature engineering tools ``\textit{are not automated enough}" and thus there is a ``\textit{need for supervision}". And, again ``\textit{automated feature engineering often comes short to domain knowledge, automating the latter is a difficult problem}". Also, as the feature engineering tasks are likely to be problem dependent and ``\textit{are mainly limited to tasks and types of data sets}", thus for tools it is ``\textit{very difficult to fine-tune manually}" for diverse problems, data, and algorithms. 

\paragraph{Domain Knowledge Requirements:}
Like other phases of ML workflow, feature engineering requires ``\textit{too much expert knowledge}" in the associated problem domain. Also, ``\textit{some problems require domain knowledge, that is difficult to translate in a general way to an open-source tool}". The requirements for domain knowledge limit the usability of the tools or methods to be used by experts only in a very specific domain. 

\paragraph{Adaptability:}
As mentioned, feature engineering for ML applications is likely to be problem and data specific. ``\textit{Sometimes (the tools are) not exactly what you (practitioners) look for}". The tools for feature engineering is thus ``\textit{limited to tasks and types of data sets}". Tools designed for a specific problem and data set or types of data may not be easily adapted to other problems or data. Feature engineering tools are not flexible enough ``\textit{to fine-tune}" for a specific problem and data sets or there are ``\textit{difficulties with setting specific properties}" to accommodate new data sets. Again, feature engineering tools are usually equipped with a static set of features and ``\textit{they do not learn, it's a fixed set of algorithms}" to exhibits robustness to diverse data and problems. As the tools are too tied to the problem domain and data types they may not ``\textit{scale, (and) guarantee performance on different platforms}". 

\paragraph{Usability:}
Another important issue with the feature engineering tools or methods is the lack of ``\textit{simplicity}". This may result in poor usability, leading to slow ``\textit{learning curve}" and may require expert knowledge in the domain. There is also a lack of ``\textit{versatility and good documentation}" to use the tools effectively. 
\paragraph{Feature Evaluation:}
It is also important but difficult to evaluate the quality or performance of the resulting features from a feature engineering tool. ``\textit{There is no integrated solution to test how the features perform with a given set of model architectures. Most of the time, we still need to do this manually (with grid search or Bayesian optimization)}". 

The practitioners also mentioned diversity in data types, adjusting too many tools or method parameters, and limitation of implementation language as some of the challenges in feature engineering. One practitioner mentioned the translation of user behaviours as an important challenge arguing that ``\textit{User behavior is inherently difficult to encode in any feature space, but it's the domain of interest for much of anomaly detection in cyber security}". The practitioner also expressed his expectation as a ML practitioner as ``\textit{I'd like to see more industry-ready research in this field}".   

It is also interesting that the perception of the limitations of the feature processing tools varies widely among the practitioners. This variation is likely due to the differences in ML application development contexts of the practitioners or might also be due to the lack of awareness of the practitioners regarding the needs and challenges in different ML development scenarios. In one hand, while we have above-listed limitations pointed out by many practitioners, some other practitioners, on the other hand, do not see any limitations of the existing tools mentioning ``\textit{None (no limitation), they are great}" or ``\textit{I didn't see real problems}" or even ``\textit{no idea}". Thus, it is important to identify domain-specific challenges in feature processing to make the practitioners aware of the challenges within and beyond their domain of ML expertise. 

\vspace{0.2cm}
\begin{tcolorbox}
\textit{\textbf{Finding 9}}: As outlined by practitioners, the common limitations of the existing feature engineering tools and methods include generalization, scalability, automation, domain knowledge requirement, and adaptability. It is also challenging to evaluate the quality of features and the performance of the feature processing tools and techniques.   
\end{tcolorbox}

\subsubsection{RQ10: What is the state-of-the-practice in feature quality assessment in ML application development?}
Once features for ML is extracted, it is important to validate the quality of the features since 
a poor feature quality is likely to affect the performance of the model negatively. The survey participants have shared their practices for assessing the feature quality. Based on the responses we observe the following common practices for feature quality assessment: 

\paragraph{Statistical Analysis and Visualization:}
One common practice mentioned by the ML practitioners for feature validation is that they apply different statistical analysis on the features. Statistical techniques include computing \textit{correlation matrix} of the feature columns, measuring \textit{mutual information}, \textit{variance}, and performing \textit{statistical tests} to understand the distribution and relationships among different features. ML practitioners also assess the feature quality ``\textit{through visualization}" of the feature. Practitioners also validate feature quality by ``\textit{estimating similarities between feature vectors to make sure they stay consistent}". 

\paragraph{Feature validation by Model Performance:}
Instead of proactive assessment of feature quality before ML model training, many practitioners rely on resulting model performance. Practitioners ``\textit{train model(s) and test them}" as ``\textit{mostly a good enough model justifies the data}". It is based on the strategy that ``\textit{we (practitioners) don't care if they (features) are representative as long as the downstream performance is good}". Commonly used model performance metrics are accuracy, precision, recall, F-score measured ``\textit{by running accuracy tests on the model with respect to the validation set}". The validation can also be done through the k-fold cross-validation of the model. Before training the models on the whole data set, one approach that practitioners use is to build a prototype model ``\textit{by running machine learning algorithms on subsets and comparing performance}" to estimate model performance and the feature quality. Another approach for feature evaluation is by ``\textit{running baseline algorithms such as Logistic Regression and see how they bare on each set of features}". Practitioners may also  ``\textit{compare performance of multiple models}" to evaluate corresponding feature sets.   

\paragraph{Feature Validation by Feature Selection:}
Another approach the practitioners reported for feature validation is that individual feature is selected incrementally (forward selection) for the model and the model performance is observed to decide on the inclusion or exclusion of the feature. Alternatively, modeling can begin with selecting all the features and then gradually eliminating (backward elimination) features based on the resulting model performance, to find the best feature subset. Given that the number of features can be high and the training process can be costly, one practitioner pointed out the limitation that ``\textit{since my own capacities are limited, I would use forward/backward selection if I can rapidly train a model}". 

\paragraph{Domain Knowledge Based Feature Validation:}
Practitioners select and validate features based on their domain knowledge. One practitioner mentioned that ``\textit{I would rely on my good judgment. Would that feature help me, as a human, make a prediction.}". Similarly, another practitioner presented the importance of domain knowledge in feature selection as ``\textit{Imagine that you are the model, and ask yourself "Am I able to predict the outcome given only these information only?" If the answer is yes, the features represent the characteristics of the dataset}". The domain knowledge can also be useful in the manual inspection of the features and the model performance. 

\paragraph{No Feature Validation:}
Depending on the domain and the types of the problem and the data, practitioners may not always need to validate features. For example, one practitioner from NLP domain mentioned ``\textit{I have used the traditional feature extraction methods and have never looked to validate them}". Also, in deep learning, explicit feature processing and validation may not always be required. 

 Besides, practitioners also adopt domain-specific techniques for validating feature quality. For example, in cases of generated (synthetic) features, one way to evaluate generated feature is to measure ``\textit{the distance of the artificially generated samples and the real distribution}". Some practitioners also reported to use ``\textit{model interpretability/explainability (such as LIME\footnote{https://github.com/marcotcr/lime}, SHAP\footnote{https://github.com/slundberg/shap})}" to assess the model and so the quality of the features used in those models. 

\begin{tcolorbox}
\textit{\textbf{Finding 10}}: The approaches practitioners commonly use for the assessment and validation of features are: statistical analysis, visualization, evaluating the resulting model performance, and incremental feature selection. Besides, practitioners also depend on knowledge and expertise in the domain to evaluate features.  
\end{tcolorbox}

\subsubsection{RQ11: What are the common practices for feature selection in ML application development?}
Once the feature is extracted from the data, one key task is to select the suitable subset of the features that best represent the data characteristics. The practitioners were asked to share their adopted practice in feature selection. As in other phases of ML application development workflow, domain knowledge plays an important role in feature selection. Around 75.5\% (37/49) of the practitioners reported that their feature selection is based on the domain knowledge. Besides, 63.26\%(31/49) of the participating ML practitioners mentioned that they do feature selection based on the statistical analysis and visualization of data and feature correlation. Also, 51.02\%(25/49) practitioners reported the use of automated feature selection tools and techniques. Among the practitioners, 40.8\%(20/49) mentioned that they take an incremental approach to add one feature at a time and evaluate the model performance for selecting that particular feature. Alternatively, all features can be added and then some of them removed gradually to find the best subset of features. It is to be noted that the above counts of practitioners are not mutually exclusive, rather practitioners use the feature selection approaches based on the specific context of the problem and the associated data. 

\vspace{0.2cm}
\begin{tcolorbox}
\textit{\textbf{Finding 11}}: The majority of the practitioners (75.5\%) use domain knowledge for feature selection. Besides, statistical analysis and visualization, using automated tools, and incremental selection of features are the common techniques adopted by the ML practitioners for feature selection. 
\end{tcolorbox}

\subsection{Model Building}
Building ML models comprises of implementation and training of the ML models. Models are trained on the training data after implementation until certain quality is achieved by the models measured against selected quality metrics. We discuss the common practices in ML model implementation and testing as follows:
\subsubsection{RQ12: What are the practices for ML model implementation commonly adopted by the practitioners?}
There are a number of popular libraries and frameworks for model implementation in different programming languages, supporting different application domains and platforms. Based on our survey responses, we observe that the implementation of ML models is primarily based on existing ML libraries and frameworks. About 93.18\%(41/44) of ML developers reported that they depend on ML libraries and frameworks for implementing ML models. About one-third (31.81\% (14/44)) of ML practitioners reportedly implement their ML model training code from scratch than relying solely on the ML libraries. Practitioners also mentioned using their own custom auto-ML system for ML model training. It is to be noted that the above distributions are not mutually exclusive and developers are likely to adopt implementation strategies that best fit their ML development contexts. 

\begin{tcolorbox}
\textit{\textbf{Finding 12}}: ML developers primarily (93.18\%) depend on the existing libraries and frameworks for implementing ML models while roughly one-third of the practitioners write code from  scratch for model implementation. The implementation choice is influenced by a specific ML development context.  
\end{tcolorbox}

\subsubsection{RQ13: What is the state-of-the-practice for ML model implementation testing by ML practitioners?}

As mentioned earlier, one of the important and challenging task for ML model implementation is to test the implementation. Given the expected challenges, we were particularly interested in knowing the state-of-practices from the ML developers; \ie{} how they validate their ML model implementation in real-world development scenarios. Based on the responses of the practitioners we identify the following practices that the ML developers adopt for testing ML model implementations:

\paragraph{Performance-based testing:}
One common practice that the practitioners follow to test ML model implementation is to evaluate the model performance. The performance is tested based on the known validation set. Practitioners test models ``\textit{mainly through measuring the performance on the test dataset, or through cross-validation}". Based on the responses of the practitioners, we observe the following practices commonly used for performance-based testing of ML models: 
\begin{itemize}
    \item \textit{Sanity checks}: Developers do some sanity checks by inference on random data samples, checking for corner-cases or overfitting the models on small sample data subset. A quick visualization of output \ie ``\textit{metrics, train/test loss curves, etc.}" can also be useful. The models can be tested on ``\textit{on simple crafted data}" or using ``\textit{small dataset} of ``\textit{known cases}".   

    \item \textit{Performance on benchmark data sets}: One approach to test ML implementation is to measure performance on some well-know benchmark data set. For example, one participant suggested the testing of ML models to be done ``\textit{by running on classical data sets such Iris or Boston and infer correctness based on the results}". Thus, domain-specific data sets can be used for the evaluation of ML model implementation. 
    
    \item \textit{Performance compared to baseline models}:
    A comparison of model performance with the known baseline model can also be used for model testing. One practitioner suggested to ``\textit{compare its accuracy to the simple baseline model that you are sure you cannot mess up. If it's better or the same then you are probably implementing it correctly...}".
    
    \item \textit{Cross-model testing}: ML developers also compare model performance with other models of different configurations to identify possible issues. Practitioners do this ``\textit{by monitoring the model parameters and also the model prediction errors}".  
    
    \item \textit{Cross-algorithm testing}: Developers can also test the model by comparing model performance with models based on different other algorithms. Practitioners ``\textit{check results compared to other methods, observe the prediction}" to verify model implementation.
    
    \item \textit{Cross-language testing}: Practitioners also reported to compare models implemented in different languages for testing. The strategy is to ``\textit{established examples, area-specific toy examples, sometimes compare with implementation in another language}" or ``\textit{compare with other libraries}".
    
    \item \textit{Cross-platform testing}: One of the strategies the ML developers take to test model implementation is to compare the model performance in different platforms or by comparing ``\textit{against one or more other frameworks}". 
\end{itemize}

\paragraph{Visualization:}
Another technique the practitioners use to test model is by using the visualization of model output (\eg{} accuracy) and other model states (\eg{} loss) or parameters. In one practitioner's word ``\textit{I have found that visualizing either the output or the internal state of a neural network, greatly improves my bug-finding capacity}". Also, another practitioner mentioned ``\textit{I can test if the neural network architecture definition is correct generating a visualization of architecture using Tensorflow, and check if it's logically correct}".  

\paragraph{Use available tools and frameworks:}
Practitioners also use features available in the existing tools and frameworks \ie ``\textit{through test methods provided with the framework}" for debugging and testing ML model implementation. Commonly it is done by ``\textit{debugging and checking results}", for example, using ``\textit{unit tests in C\#}". An example approach as mentioned by one ML practitioner is ``\textit{in case of python, I will use PDB(python debugger). Then test it over the known outcomes and if they are not correct or efficient and I will check the algorithm I am using and the outputs at every step of that model}". Some practitioners write suitable unit-tests for ML models or test ``\textit{against existing frameworks}". Some practitioners use interactive interface of Jupyter notebook and examine ``\textit{incremental results from the different code blocks}". 

\paragraph{Domain knowledge based validation:}
Similar to other activities of ML application development, domain knowledge plays an important role in testing ML code. 
Based on the domain expertise, practitioners can perform visual or manual inspection of model behavior based on known cases or crafted test data. 

\begin{tcolorbox}
\textit{\textbf{Finding 13}}: 
The majority of the developers depend on the existing ML libraries and frameworks for implementing ML models than writing code from scratch. Practitioners test the model implementation by observing performance, visualization of model structure or the outcomes, testing or debugging code using existing tools. Domain knowledge plays an important role in testing ML implementation.   
\end{tcolorbox}


\subsection{RQ14: What are the common symptoms that practitioners use to detect defects in an ML implementation?}

It is known that the identification of defects in ML code is harder because the outcomes of ML applications are generally stochastic in nature. Thus we are interested to know how ML practitioners detect defects in their ML code. The shared knowledge of the practitioners will not only be useful to the ML community but also help in identifying the challenges practitioners face and the types of support they need for defect identification in ML code. The symptoms can be intuitive like ``\textit{fail to compile}" to some nontrivial defects symptoms. Based on the practitioners' experience shared in this survey, we observe the following practices commonly used by the ML developers to identify defect's symptoms:  

\paragraph{Performance based symptoms:}
ML practitioners emphasize a lot on the model behaviour or performance as the defect symptoms for ML models. ML developers frequently consider the following performance-based symptoms for defects in ML code. 
\begin{itemize}
\item{Accuracy:}
In the practitioners' view, model accuracy is a strong indicator of the correctness of the implementation. In  a practitioner's word,  ``\textit{Highly unlikely accuracy for the given task, and extremely low accuracy for the given task, often when encountering these results there would be defects in the implementation whether it is minor or major}". So, the ``\textit{weird results}" \ie~``\textit{extremes, too high accuracy, too low accuracy} is considered to be an indicator of defects. Also, an abrupt change in model accuracy such as ``\textit{huge decrease in performance}" is also a sign of model defects.  

\item{Consistency:}
Inconsistency in model performance is likely to be a symptom of defects in the ML model. The ``\textit{inconsistency of results, over-sensitivity}" \ie~``\textit{non-deterministic results}" indicate the likely presence of defects in the ML code. 

\item{Generalization:}
Another symptom of defects in ML implementation is the poor generalization \ie~model exhibits ``\textit{decay in performance on unseen data}". Models may have ``\textit{wrong output,(and) limited possibilities to generalize}" and may also show ``\textit{discrepancies between offline and online benchmarks}" in presence of defects. 

\item{Bias:}
ML models may also exhibit ``\textit{high bias with respect to the labels}" in presence of defects. Such ``\textit{high bias or high precision with low bias}" might be an indication of defects in the model. This bias can originate from both data and the code. 
\end{itemize}

\paragraph{Training behaviors:}
Some symptoms of defects can surface during model training. We list some of the defects symptoms observed during training of ML models as follows:
\begin{itemize}

\item {Convergence:}
In presence of defects, ML models may fail to converge. The model may lead to ``\textit{overfitting, underfitting, and volatile performance}".

\item {Speed:}
The models may be too slow in training and inference or can be too fast with high accuracies. These unexpected model behaviours can indicate potential defects. 

\item {Value:}
Models may have unexpected values (types or range of values) for input, weights or parameters during training. For example, one indication can be ``\textit{appearance of NaNs during training}". 
This can indicate potential defects in the model implementation. 
\end{itemize}

\paragraph{Model output:}
Erroneous model output may also indicate presence of defects. Practitioners reported the following defect symptoms related to model output:
\begin{itemize}
\item {Value:}
Model can produce wrong output in terms of values and range of values indicating defects in the model.  
\item {Distribution:}
The distribution of model output can also be an indicator of the model defects. If the model output is skewed to some specific class or values, there might be defects or bias in the implemented model. 
\end{itemize}
However, the symptoms of defects in ML code may also ``\textit{depend on the problem}" and can be hard to fit in specific symptoms. The intrinsic challenges in identifying defects in ML code have also been reflected in some of the practitioners' lack of awareness of the identified symptoms; mentioning that they have ``\textit{no idea}", or they are ``\textit{not sure}" or, that defects symptoms are ``\textit{unknown}". This shows the importance of providing ML practitioners with more (tool) support to help them detect defects in their code. 


Practitioners use different tools and techniques for detecting bugs in ML code.  Some practitioners reported that they use Pytest for testing ML code in Python. Based on the ML frameworks, the test techniques vary widely. Some practitioners simply use logging or debugging to identify bugs. Interestingly, many practitioners either do not use any particular tool for ML testing while some practitioners are not even aware of such events. Some other tools and frameworks such as: Python debugger available with the IDE like PyCharm can be handy for looking for bugs. 

\vspace{0.2cm}
\begin{tcolorbox}
\textit{\textbf{Finding 14}}: Extreme (too good or too bad) model performance, inconsistency, generalizability, and bias in model outcomes are the common symptoms of defects in ML code. Besides, poor model convergence, unusual training and inference time, and output values and their distribution can be important indicators of the presence of defects. 
\end{tcolorbox}



\subsubsection{RQ15: What are the practitioners perceived challenges of testing ML application?}
Testing the correctness of models is a challenging task. The characteristics of the ML Models (algorithms) and different quality requirements of ML data make the testing of ML models more difficult. Based on the developers' perception of the challenges in ML model testing, we list the following challenges in ML model testing: 

\paragraph{Black-box nature of ML models:}
ML models are ``\textit{black-box in nature}" \cite{Pei_SOSP_2017} meaning that it is quite obscure how the model perform a particular task. For the same reason, it is also hard to detect or explain why a model is not performing as expected. As models are black-box ``\textit{I (developer) can't tweak the internals}" to identify the issues in the model. Again, ``\textit{since they are mostly black boxes, it's hard to make guarantees on yet unseen data}". The opaque nature of the models usually does not allow the developer to observe the internal states of the models such as ``\textit{gradient inspection}" during training. The practitioners feel the need for techniques or tools to make the model more transparent. In practitioner's word ``\textit{... it would be awesome some kind of software or library that checks your gradients and tell if there is something strange}". 

\paragraph{Model's robustness to errors:}
Another key challenge to test ML model is that ML models can exhibit robustness to errors by producing correct results in some cases \ie{} ``\textit{the model is wrong, but the result is not}". Similarly, ``\textit{a wrong implementation sometimes achieves similar performance (to the correct ones), and the bug cannot be found until we introduce new features}". Such wrong implementation ``\textit{without actual effect in results while convincing myself(developer) that it works}" can be quite tricky for the ML developers to identify and fix. 

\paragraph{Data Quality:}
Data plays an important role in model quality. Despite the correct implementation, model may perform poorly due to issues in the dataset and that can be hard to detect. Based on the practitioners' responses, we observe the following challenges in testing ML models related to the quality of data:
\begin{itemize}
    \item Adequacy: ML models need to be trained on an adequate number of input data to achieve high accuracy. ``\textit{Absence  of sufficient correctly labeled data}" can hinder the correctness of the model and consequently the performance.  
    
    \item Correctness: One key concern in ensuring the correctness of ML models is ``\textit{having a bad model due to bad data: makes it very difficult to detect and expensive to solve}". 
    
    \item Data Bias: Biases in the data can lead to poor quality ML models. There might be intentional or unintentional or even domain-specific biases in the data set. For example, ``\textit{in the medical industry, data bias is hard to get over, and typically requires vast amounts of augmentation}". Unbalanced distribution of data samples in the training, testing and validation data set can embed bias in the model leading to biased models. 
    
    \item Labeling accuracy: Incorrectly labelled data is likely to result in poor quality models. Thus it is important to ensure ``\textit{the fidelity of the labels}" because the ``\textit{lack of (a) good labelled data sets}" may adversely affect the model accuracies.   
    
    \item Distribution: ``\textit{Testing data sometimes does not belong to the distribution of the training data on which the model is trained so we might not get good score}". ML data also need ``\textit{to represent the real distributions faithfully}". Otherwise, this can lead to incorrect models which can be hard to identify and fix. 
    
    \item Divergence: One important challenge for the ML model is that the characteristics of the data can evolve over time. Thus, the issues in the models may be simply ``\textit{a data problem}". For the models to be correct, it is important ``\textit{how well does the data available during training reflect true operational data, and how long until the nature of the operational data diverges from your original data and assumptions}". It is challenging to deal with this divergence in ML data. 
\end{itemize}

\paragraph{Volatile performance:} The performance of the ML models can be affected by diverse factors involving both data and the code. ``\textit{Sometimes there are discrepancies between test and validation sets and unseen data}". Again ``\textit{sometimes the models do not meet the predetermined correctness criteria or they perform very well in the train test split but when introduced to the validation set it under-performs}". The cause of low performance can be challenging to identify. 

\paragraph{Domain Expertise:}
Testing ML models may require adequate domain knowledge. Thus, ``\textit{lack of domain knowledge}" can make ML testing a difficult task. Also, another challenge is the availability of domain experts and also to ensure the availability and ``\textit{access to tools a domain expert can use}".  

\paragraph{Cost:} Testing ML models for correctness and performance can be costly regarding the time and efforts. Phases of the ML application development are usually iterative and thus ``\textit{reaching a working model is very time consuming}" because ``\textit{iteration takes too much time}". 

\paragraph {Lack of concrete methodology:}
Practitioners are also in lack of appropriate techniques and methodologies to test ML models. In a developer's view ``\textit{... there is, to my knowledge, no decisive way to ensure correctness but to leverage more data for testing predictions}". ``\textit{You (practitioners) only have accuracy and prediction plot to check the correctness. Usually, the data transformation pipeline is difficult to implement so the bug can be from data, even before the model. So you(practitioners) need to identify exactly where is the cause of low accuracy}".  

\paragraph{Interpretability or explainability:}
Another key challenge in testing ML models is that we can merely explain why and how a model is working or why it is not working. To test ML models, it is hard to ``\textit{knowing what you (developers) are evaluating}" \ie ``\textit{understanding the exact mathematical process behind the algorithms}".    

\begin{tcolorbox}
\textit{\textbf{Finding 15}}: The black-box nature of the ML models, robustness to errors, and the characteristics of the data including adequacy, correctness, bias, labelling accuracy, distribution and divergence of the data pose key challenges in testing ML applications. Diverse performance impacting factors and the lack of concrete testing techniques also pose challenges in ML testing. Requirements for domain expertise, time and efforts and lack of explainability of ML models also hinder ML testing. 
\end{tcolorbox}

\subsection{Model Deployment and Management}
Once trained and tested ML models are available, models need to be integrated into the target application for deployment. Also, deployed models need to be monitored and manged to maintain the expected performance of the application, and adapted to changes over time. 



\subsubsection{RQ16: What are the developers-perceived challenges of testing ML model deployment?}
It is important to test models after deployment to ensure that the model is integrated as expected with other components of the ML applications and the target software ecosystem. However, like pre-deployment testing of models, there are some challenges to post-deployment testing. Based on the perceptions shared by the survey participants, we list the following challenges for testing ML model deployment:
\paragraph{Test Data:}
Quality of test data is an important challenge to test ML model deployment. Practitioners identify the following challenges in testing model deployment:
\begin{itemize}
    \item Data Availability: For post-deployment testing, adequate real data needs to be available covering possible usage scenarios including corner cases. However, there might be a ``\textit{lack of ground truth for real life new data}". While in some cases it might be challenging to have adequate test data, ``\textit{huge amount of (test) data}" can also be a challenge to deal with. 
    \item Data labelling: As in the training phase, test data requires correct labelling for post-deployment testing. However, ``\textit{obtaining correctly labelled data}" can be challenging.   
    \item Data format: Besides the availability of correctly labelled data, another challenge is  ``\textit{making sure the data coming into the model is of the right format}".  
\end{itemize}
\paragraph{Performance:}
One key requirement is that the deployed ML models must perform as expected in production. The practitioners pointed out the following challenges to ensure the performance of the deployed model:
\begin{itemize}
    \item Functional accuracy: One of the primary requirements that a deployed model must satisfy is the desired level of functional accuracy. The accuracy requirement may ``\textit{depend(s) on applications, some applications require very high accuracy, while for others 60\% is enough}". 
    
    \item Generalizability: Another challenge in post-deployment model testing is to ``\textit{making sure the model is generalizable and its behavior is in control}". The deployed model expected to be not only generalizable to unseen data, but also exhibit ``\textit{robustness to adversarial examples}". 
    
    \item Performance monitoring: The deployed models need to be evaluated over a reasonable period for post-deployment testing. However, ``\textit{keeping track of predictions (model performance)}" can have additional overhead. Also, ``\textit{constantly having to have human oversight}" can be challenging and costly. 
    
    \item Performance measures: Sometimes the interpretation of the performance metrics may differ and ``\textit{the most appropriate result metrics may or may not be understood by all the teams}". For example, ``\textit{accuracy they understand and F1-Score is what matters}". 
    
\end{itemize}

\paragraph{Resource requirements:}
The deployed ML application should have optimum requirements for resources such as processing powers and memories. The models should be tested against these requirements. 
``\textit{Assessing the accuracy is not hard but It's hard to measure resource usage in mobile phones. It's difficult because in the phone we don't have the same libraries and tools we can use to develop like on PC}". 

\paragraph{System Complexity:}
Complexities of the model, target application architecture and the deployment environment can pose challenges to testing deployed models. For example, ``\textit{due the model complexity, it is hard to understand where the problem is from}" and to devise test cases for all possible scenarios.  

\paragraph{Platform diversity:}
Another important challenge in post-deployment testing of ML application is that model development or training platforms can be different. For example ``\textit{deployment hardware is different from (the) hardware used for training}". Models trained and tested on different hardware and software environments may not work on a mobile platform. And ``\textit{it's difficult because in the phone we don't have the same libraries and tools we can use to develop like on PC}". ``\textit{The big challenge is a method to deploy ML models through a single framework regardless of the used library (PyTorch, Tensorflow, etc.)}" to overcome the complexities due to platform differences. 

\paragraph{Adaptability:}
The target environment and data characteristics are likely to evolve over time. However, it is challenging to ``\textit{integrating new cases of failure into the pipeline}" to ensure the adaptability of the deployed ML application.  

\paragraph{User satisfaction:}
The success of the deployment does not include satisfying the functional or performance requirements but also how it is satisfying the target users. However, challenges remain in ``\textit{determining the value of one set of (the) user over another. If 60\% of users dislike the current implementation, should we change it to satisfy their needs but put the other 40\% in a place of discomfort?}". The deployment testing thus should consider the user acceptance of the application or model deployed. 

Besides, the practitioners claim that domain knowledge requirements, associated time and costs, complexities in writing suitable tests, and lack of interpretability or explainability can also pose challenges in post-deployment testing as in other phases of ML application development. 

\begin{tcolorbox}
\textit{\textbf{Finding 16}}: 
The key challenges in testing the deployment of ML models are associated with the data quality (availability, format and labelling accuracy) and the model performance evaluation (functional accuracy, generalizability, performance tracking, and metrics). Moreover, model complexity, resource requirements, target platform diversity, adaptability and overall user acceptance pose challenges to post-deployment evaluation and testing of ML applications.  
\end{tcolorbox}

\subsubsection{RQ17: What are the factors that ML developers commonly focus on during ML model management?}

Post-deployment model maintenance is important in the ML application development life cycle. In this phase, models need to be monitored for different quality parameters and model maintenance activities need to be initiated if the model deviates from the performance requirements. About 77.5\% practitioners have mentioned that they frequently monitor ML models after deployment. ML developers employ different testing techniques depending on the context of the specific application. Common model maintenance activities include observing the performance, resource requirements, and robustness of the models in real-world usage scenarios. The models are deployed on different software and hardware platforms. For example, the models can be hosted in some local server or can be deployed on the cloud. These diversities in the deployment environment is  likely to have an impact not only on the performance but also on the maintenance of the models. 

Monitoring and maintenance of deployed models are essential in ML application maintenance. Since ML models are data-driven, deployed models need to be monitored because the model performance can be affected due to significant changes in the data over time. To have insights into the practices that ML developers commonly adopt in managing ML applications, we categorize the common parameters considered by the practitioners for post-deployment model management as follows: 

\paragraph{Performance:}
Monitoring performance is a key task in ML model management. As model performance can drop significantly due to changes in data characteristics, models need to be monitored to detect performance deviation. As per the practitioners' responses, the following performance factors are considered for post deployment monitoring.

\begin{itemize}
    \item Model accuracy: One key performance measure is accuracy. In the maintenance phase, models need to be monitored for model accuracies. This helps in the detection of performance deterioration due to changes in data over time. Model accuracies are measured against predefined metrics similar to the training and testing phases. ``\textit{Changing in accuracy}" should be addressed accordingly such as retraining the models. The metrics for accuracy may vary depending on the model type and the domains. Models are also evaluated with respect to the ranking quality in recommendations such as ``\textit{NDCG (Normalized Discounted Cumulative Gain)}" commonly used in information retrieval.  
    
    \item Resource consumption: 
    Another performance factor that the practitioners reported to give importance is the resource consumption by the deployed application such as CPU, GPU, memory, and power. The models may require optimization for resource consumption to ensure cost-performance trade-off. 
    
    \item Inference latency: Response time for inference by the models \ie{} the inference latency is an important parameter to watch for. Models require to be evaluated for the ``\textit{latency for predictions}" to ensure that ``\textit{speed is reasonable}". 
    
    \item Robustness: The models need to be monitored for unseen data or corner cases to evaluate how robust the model is in dealing with new data in a real environment. 
\end{itemize}
\paragraph{Business gains:}
In the post-deployment phase, ML models or applications also need to be evaluated regarding different business metrics. Thus, ML applications are evaluated regarding different ``\textit{business metrics such as click-through rates, conversion rates, (and) revenues}". There are also other business factors such as ``\textit{customer retention}". Necessary steps are essentials if the ML model fails to meet the business goals. 

\paragraph{User Feedback:}
It is also important to evaluate ML applications based on the feedback of real users. ``\textit{Response from users and their feedback in how it could be improved or changed}" using the ``\textit{end-user perspective}" is important to improve the quality of ML applications. 

However, the prioritization of factors to monitor during model maintenance can depend on the associated ML applications. 

\begin{tcolorbox}
\textit{\textbf{Finding 17}}: 
 From practitioners' view, the factors that need to be monitored for deployed models are primarily the performance (accuracy, latency, robustness and resource consumption), business parameters (\eg~revenues, customer conversion or retention rate, click-through rates) and the overall user acceptance of the ML applications.  
\end{tcolorbox}


\section{Analysis and Discussion} \label{sec:discussion}
In this paper, we have presented insights into state-of-the-practice and key challenges in different phases of ML application development based on the shared experiences from ML practitioners. Our survey participants are from  diverse backgrounds with a wide array of skills and experience in different domains of ML application development. This is likely to ensure comprehensive reflections of the practices and challenges in machine learning in practice. 
In the following subsections we discuss our findings with respect to overall trends in ML application development and the practices and challenges in the four phases of the ML application development life cycle we covered in our survey:

\subsection{Trends in ML Application Development}
We presented recent trends in ML application development in \textit{Finding 1}. As mentioned by the practitioners, the recent trend in developing ML applications is heavily focused on developing Business Intelligence (BI) applications. In addition to business and e-commerce, the application of AI/ML includes healthcare, security, document analysis, entertainment, and embracing rapidly other areas of human life. For ML application development, data plays a key role. Based on the practices reported by the practitioners, open-source is the leading source of ML data while private company data and data from third-party clients are also major sources of ML data. Like open-source dataset, different open-source ML libraries and frameworks (\eg~TensorFlow, PyTorch, Keras, scikit-learn) are the leading frameworks as reflected in the recent trends in ML application development. However, the choice of the data, ML algorithms, and ML frameworks are mostly dependant upon the problems and application domains and require necessary domain knowledge for the successful development of ML applications. 

\subsection{Data Collection and Preprocessing}
ML models are data-driven and thus the quality and adequacy of data are important for developing ML applications. However, ensuring the availability of reliable data for ML can be challenging. 
We observe that the open-source data is the most prevalent source for machine learning data covering about 75\% of the data. 
The quality of ML data is very important for the performance of the resulting ML model. Practitioners have pointed out different key quality characteristics of the ML data (\textit{Finding 2}). The key attributes of ML data that the practitioners have emphasized on include how well the data represent feature for machine learning and also the adequacy and diversity; meaning enough data volume and having representations from all class or categories. ML data also needs to be complete and accurately labelled for ML algorithms. Data needs to be consistent regarding the structure, accuracy, and quality of information. Besides, ML data needs to be reliable, possibly verified in multiple phases because the consequence of data error can be extremely adverse in cases such as health- and safety-critical systems. Besides, ML data should have if possible low noise and bias, with balanced distribution across data classes, to achieve high performance models. 

Practitioners apply different data transformation operations such as noise removal, replacement of missing values, dimensionality reduction, class-balancing and normalization (\textit{Finding 3}). As we observed from the practitioners responses, there is no one-size-fits-all type of solution for data processing. More than two-third (76.6\%) of the surveyed developers do not use specific data analysis tool, rather, they use diverse data or problem specific tools and techniques or develop their own customized solutions. 

ML data can be noisy and may require preprocessing such as cleaning or transformations to be suitable for ML models. However, cleaning ML data is a challenging task (\textit{Finding 4}) and the data cleaning approaches are likely to be data and problem-specific. Thus, the approaches are hard to generalize, scale up to accommodate large volume of data, and to apply on data automatically. Quality issues of the data and lack of necessary features in existing tools for ML data processing make the data cleaning task not only harder but also costly in terms of the time and efforts. The process requires adequate domain knowledge about the associated data and the ML use-cases. Practitioners also pointed out the importance of having a common standard for data and data cleaning procedures. It is also important for the practitioners to be aware of not only the tools and techniques for data cleaning but also the adverse impact of noisy data on resulting ML models. 

Another key task is to correctly label the data or features for ML models. Practitioners have outlined several challenges in feature labelling (\textit{Finding 5}). One key challenge in feature labelling is the large volume of data which is time and resource consuming and thus can be costly as it may also involve manual processing. The feature extraction and labelling procedures are likely to be problem and domain-specific and thus domain knowledge is important. Poor quality of data and lack of appropriate labelling guidelines also add challenges in ML feature labelling. As reported by the practitioners, manual investigation is still the most commonly used approach for validating feature or ML data labelling where domain knowledge plays an important role (Finding 6). However, some practitioners use tools and automated scripts for feature validation.    

\subsection{Feature Engineering}
Feature engineering is another important phase of ML application development where ML data is processed to generate meaningful features. The key objective of the features is to best represent the data characteristics that can help ML models to learn and infer for a defined ML task. Feature engineering usually comprises of two key functionalities: feature extraction and feature selection. The feature extraction process should take into account the quality characteristics while extracting features from the data. 

One important requirement for ML data is that data need to be balanced across different classes (\textit{Finding 7}). Practitioners commonly uses different statistical analysis, visualization and manual verification to check class balancing of the ML data set. To improve model performance,  practitioners fix class balancing issues in ML data using techniques such as re-sampling of data or stratification of the ML data set. As the feature extraction procedure may be data and problem dependant, more than two-third (76.6\%) of practitioners mentioned that they do not use specific tool for ML data processing such as feature extraction. Practitioners usually depend on manual analysis and custom scripting for feature extraction, while there are some tools available for feature engineering (\textit{Finding 8}). Based on practitioners' perception, we observe several limitations of existing feature engineering tools and techniques (\textit{Finding 9}). Generalization is one of the key limitations of existing feature engineering tools and associated domain knowledge is necessary to use these tools. They also are not easily adaptable to new data and problem. Practitioners also identify the lack of usability and simplicity of the tools, which imposes a slow learning curve to developers. 

It is also important to validate the feature quality. Practitioners have reported some common practices in feature assessment and validation (\textit{Finding 10}). One common practice is to use statistical analysis and visualization of the feature to evaluate the quality characteristics. Practitioners also observe resulting model performance to assess the feature quality. Like other ML development phases, feature validation requires necessary domain knowledge. However, the techniques are likely to be domain-specific. 

Once features are extracted from ML data, selecting the optimal subset of features for ML models is an important but challenging task. As per the practitioners' responses (\textit{Finding 11}), the feature selection process are mostly manual and based on domain knowledge while there are some automated tools. However, different statistical analysis and visualization are useful to gain insights into the features for incremental selection, to find the optimal subset of features that satisfy desired performance requirements.

\subsection{Model Building and Testing}
ML models are usually (reported by 93.18\% practitioners) implemented base on available libraries and frameworks while about one-third of the developers write models from the scratch (\textit{Finding 12}). The models are then trained with the training dataset and tested for accuracy and performance. One important and challenging task of the model building phase is to ensure the accuracy of the models by testing (\textit{Finding 13}). A widely used practice for ML model testing is based on model performance evaluating selected performance metrics on benchmark or validation data set. The testing may involve comparison of models considering diverse settings (language, algorithms, target platforms, etc). ML model implementations are also tested using visualization of internal model states or external behaviours. There are some tools and frameworks to support ML implementation. Domain knowledge also plays an important role in ML testing. 

Practitioners also pointed out some defect symptoms commonly used to assess the quality of ML models. One key parameter is to evaluate model performance regarding accuracy, consistency, bias, and generalizability of models. Training-time behaviors like convergence, training time and trend, and also output values and distribution from model can be useful indicators of defects in the model (\textit{Finding 14}). 

Testing of ML models is known as a challenging problem. 
Practitioners have identified (\textit{Finding 15}) that the ``black-box'' nature of the ML models makes it harder to test. ML models can also exhibit robustness to errors; meaning that ML models can produce correct results in some cases despite incorrect implementation. In addition, it is also challenging to ensure adequacy, consistency of data. Possible bias, labelling errors, and divergence in the data set also pose challenges to ML testing. Practitioners are in need of concrete methodologies for ML testing. Lack of interpretability or explainability also hinder ML model testing. 

\subsection{Model Deployment and Maintenance}
Once trained and tested, 
the model needs to be integrated into the target application for deployment. Deployment of ML models include different activities to integrate and test model in the target application environment. Practitioners adopt different tools and techniques for model deployment and post-deployment maintenance of ML models. However, model deployment involves different challenges (\textit{Finding 16}). The very first challenge is to ensure the availability of test data with desired quality and diversity to cover all use case scenarios. Monitoring post-deployment model performance is another challenge to ensure functional accuracy and generalizability of models. Meeting resource requirements, complexity of models, platform diversity, adaptability and overall user satisfaction are also important model attributes for post-deployment evaluation of ML models or applications. 

Practitioners focus on some important maintenance factors for model maintenance (\textit{Finding 17}). Over 77\% of the practitioners have reported that they frequently monitor deployed ML applications. In post-deployment phase, practitioners primarily focus on the model accuracies, resource consumption, inference latency (speed) and robustness to unseen data. Besides, different business factors (\eg~user conversion rate, revenues) are important to monitor during model maintenance. Moreover, the feed back of the target users is very important to measure the application performance from end-user perspective. Users' feed back is important for the identification of defects and the improvement of the features of the ML applications. During the maintenance, the practitioners may need to set priorities among these factors based on the specific context of the ML application. 

The findings from our study highlight valuable insights into the practice and challenges of different phases of the ML application life cycle. The findings are expected to be useful to make practitioners aware of the challenges in ML application development. We hope that they will serve as guide to help developers 
adopt best practices for developing high quality ML applications. 




\section{Threats to Validity} \label{sec:threats}
In this section we discuss some potential threats to the validity of the methodology and findings of our study. 

\textit{Threats to construct validity}: Survey is a well-known method to collect information from relevant people on a specific topic that allows us to summarize, compare and explain the knowledge and perception of the respondents on the topic of interest \cite{Fink_2003}. Thus, we adopted survey as our methodology to ask ML practitioners about their experiences in ML domains. We followed formal guidelines to design and conduct the survey and analyze the responses to have insights into the practices and challenges in developing ML applications. 

\textit{Threats to internal validity}: One important threat to the internal validity is the potential biases in responses from the participants of the survey. 
We did a pilot study to get feedback from several survey participants on the questionnaire. We refined our questionnaire and adapted the recommendation we received from the pilot study to design the final survey questionnaire. 

\textit{Threats to external validity}: Although we selected a large group of participants from diverse backgrounds and skills for this study, this group of participants may not be representative of the general population of ML practitioners. This is a potential threat to the generalisability of our findings. To mitigate this threat, we carefully selected the participants based on the their professional profiles in LinkedIn and their contribution in machine learning projects in GitHub. Also, we took care to provide many open-ended questions in our questionnaire to allow participants to express their responses with freedom. We also ensured that every question includes 
the option "Not Applicable" or "Other", to allow participants to respond appropriately if a question did not apply to them or if the respondents were not comfortable with answering a particular question(s). The open-ended questions also allowed respondents to explicitly add any other practices and challenges that they are aware of. Nevertheless, it is desirable that future studies replicate this work with more ML professionals from diverse backgrounds.

\textit{Threats to conclusion validity}: Our results from the survey is not affected by the choice of methodology used in our study. We have used descriptive statistics, simple calculation and comparisons that are likely to be independent of the analysis tools and techniques. However, a different set of respondents may result in some variations in the results. We carefully selected the respondents based on their profile and contribution and we cross-validated our data analysis and reporting methodology with at least two members of the team conducting this study. 

\textit{Threats to reliability validity}: To ensure reproducibility of our findings, our data and results are available at an online Appendix \cite{replication_package_2020}. We elaborated our details methodology for selection of the respondents, data collection and analysis. As the professional network is continuously evolving and the query results for specific keywords are likely to vary, and thus the list of participants likely to differ in future replication of the study. 

\section{Related Works}\label{sec:relatedworks}
Recent advancements in machine learning 
are making ML increasingly popular to devise innovative solutions for diverse problems. 
However, the increasing adoption of machine learning into software applications is posing additional challenges to the software development process \cite{ZhangT_2019}. 
Challenges in the traditional software engineering process have been widely addressed by researchers \cite{Sandberg_ICSE_SEIP_2017}. 
However, there is a growing need for guidelines and best practices for developing ML applications.  

Schelter \etal \cite{Schelter_IEEE_bulletin_2018} focused on ML model management regarding use cases from conceptual, data management, and engineering perspectives. Amershi \etal \cite{Amershi_ICSE_2019} highlighted challenges in AI application development at Microsoft and shared how the teams address those challenges. 
Zinkevich \cite{Zinkevich_MLRules_2014} presented guidelines for best practices in ML engineering. 
There are also guidelines for responsible AI practices \cite{Google_responsibleAI, Kriens_2019}.  
However, these guidelines are not focused to find a fit for ML into traditional software engineering process. 
Many existing literature focus on different aspects of machine learning such as data acquisition, data preprocessing, 
feature extraction 
\cite{Storcheus_NIPS_2015}, model management \cite{Schelter_IEEE_bulletin_2018}, testing \cite{Pei_SOSP_2017,Grosse_and_Duvenaud_2014, Ma_2019,Ma_ASE_2018,Ma_ISSRE_2018,ZhangTest_2019,Sun_ASE_2018} and 
deployment \cite{Schelter_IEEE_bulletin_2018, Renggli_2019,Guo_2019} of ML applications. 

Since testing is one of the most important phases of the development of machine learning applications, some studies focused on the challenges of testing Machine learning systems. Ben Braiek and Khomh \cite{HoussemBenBraiek_2018} present challenges that should be addressed when testing ML programs. 
In this paper, we report about techniques and tools currently used by practitioners to cope with these challenges. 
Song Huang \etal \cite{Huang_IJPE_2018} investigate the characteristics of Naive Bayesian classifier and DNN classifier and analyze the testing challenges of machine learning applications like: Generating reliable test oracles, Generating effective corner cases, Improving test coverage and Testing the ML applications with millions of parameters. Then some initial techniques were suggested for machine learning applications which use Naive Bayesian classifier and DNN classifier to mitigate these challenges. The other study \cite{Marijan_2019} focuses on the most prominent challenges of testing ML-based systems (Absence of Test Oracles, Large Input Space and White Box Testing Requires High Test Effort) from the quality assurance perspective, rather than model performance perspective. Then, some existing approaches which alleviate these challenges are reviewed and discussed regarding their limitations.

Few researches can be found about the difficulties faced by software developers while developing ML applications or using ML libraries. Considering the 3,243 highly-rated Q\&A posts related to ten ML libraries from Stack Overflow and classifying these questions into seven typical stages of an ML pipeline, Islam \etal \cite{Islam_2019} performed an analysis from four perspectives to understand the problems with ML libraries usages: finding the most difficult ML stage, understanding the nature of problems, nature of libraries and studying whether the difficulties stayed consistent over time. Bangash \etal \cite{Bangash_2019} studied 28,010 machine learning posts from Stack Overflow and employed topic modeling to identify key areas of interest to developers. They report that topics related to Algorithms, Classification, and Training datasets categories are frequently discussed by developers. 
Nguyen-Duc \etal \cite{Nguyen-Duc_2020} in their survey explored different contextual factors in ML application development to leverage opportunities in business. 
Washizaki \etal \cite{Washizaki_IWESEP_2019} report about a systematic literature review of both academic
and gray literatures that aimed to collect software engineering good and bad design patterns for ML application systems and software. They provide a list of software design patterns and anti-patterns that practitioners can use to improve the quality of their ML applications.

The existing literature have provided useful insights on challenges for machine learning and software engineering development mostly separately. In this paper we reconcile these two themes and 
report about challenges and best practices of machine learning application development, using 
insights from experienced ML developers with diverse expertise and application domain. 

\section{Conclusion}\label{sec:conclusion}
In this paper, we presented the findings of a survey of 80 ML practitioners from diverse backgrounds. Our survey covers four key phases of the ML application development life cycle, \ie{} (i) data collection and preprocessing, (ii) feature engineering, (iii) model building and testing, and (iv) integration, deployment and monitoring, to identify challenges and practices from practitioners’ perspective. %
We summarized the knowledge shared by these practitioners in 17 key findings. 
We believe that our findings can be useful in making ML practitioners of all experience levels in academia and industry aware of diverse challenges in ML application development. In addition, our findings can provide the practitioners with necessary guidelines and examples of best practices to adopt in their ML workflow in a context-specific way. 

\begin{acknowledgements}
We express our gratitude to NSERC and FRQ funding agencies. Our heartiest thanks to the anonymous participants for their valuable time and thoughtful responses to our survey questionnaire. 
\end{acknowledgements}

\bibliographystyle{spphys}
\bibliography{ml-bib}

\end{document}